\documentclass[
aps,
prd,
preprintnumbers,
amsmath,
amssymb,
nofootinbib,
preprintnumbers
]{revtex4-2}
\usepackage{here}
\usepackage{footnote}
\usepackage{comment,braket}
\usepackage{epsf}
\usepackage{amsmath}
\usepackage{graphics}
\usepackage{amsfonts}
\usepackage{amssymb}
\usepackage{latexsym}
\usepackage{color}
\usepackage{natbib}
\usepackage{graphicx}
\usepackage{hyperref}
\usepackage{array,multirow,mathtools}

\usepackage[svgnames]{xcolor}
\definecolor{phthaloblue}{rgb}{0.0, 0.06, 0.54}
\hypersetup{
    colorlinks=true,
    linkcolor=blue,
    citecolor=blue,
    filecolor=blue,
    urlcolor=blue,
    }

\begin{document}
\title{Greybody Factors Imprinted on Black Hole Ringdowns.\\II. Merging Binary Black Holes}
\author{Kazumasa Okabayashi$^{1}$}
\email{kazumasa.okabayashi@yukawa.kyoto-u.ac.jp}
\author{Naritaka Oshita$^{1,2,3}$}
\email{naritaka.oshita@yukawa.kyoto-u.ac.jp}
\affiliation{$^{1}$Center for Gravitational Physics, Yukawa Institute for Theoretical Physics,
Kyoto University, Kitashirakawa Oiwakecho, Sakyo-ku, Kyoto 606-8502, Japan}
\affiliation{$^{2}$The Hakubi Center for Advanced Research, Kyoto University,
Yoshida Ushinomiyacho, Sakyo-ku, Kyoto 606-8501, Japan}
\affiliation{$^{3}$RIKEN iTHEMS, Wako, Saitama, 351-0198, Japan}
\preprint{YITP-24-35, RIKEN-iTHEMS-Report-24}

\begin{abstract}
The spectral amplitude of the merger-ringdown gravitational wave (GW) emitted by a comparable mass-ratio black hole merger is modeled by the greybody factor of the remnant black hole. We also include the post-Newtonian correction to the greybody factor model. Our model includes only a few fitting parameters, which could evade the overfitting issue. 
We perform the mass-spin inference from the SXS data without tuning the data range of each SXS waveform. Also, we find that the exponential damping in the ringdown spectral amplitude can be modeled well with the exponential damping in the greybody factor at high frequencies. Our findings could be consistent with a conjecture that the light ring of the remnant black hole, which sources the ringdown, forms as early as during the merger stage. We discuss the formation of the light ring in the static binary solution as a first step towards the understanding of how the separation of merging black holes may affect the formation of the light ring.
\end{abstract}

\maketitle

\section{Introduction}
Modeling of gravitational-wave (GW) ringdown has been actively studied so far as it is important to test gravity and to probe new physics in strong gravity regimes.
The most standard ringdown model is the superposed multiple quasinormal (QN) modes. Each QN mode has a complex frequency $\omega_{lmn} = f_{lmn} - i (\tau_{lmn})^{-1}$, where $f_{lm}$ and $\tau_{lm}$ are real values and the subscripts $(l,m,n)$ represent the multipole mode $(l,m)$ and the overtone number $n$. The complex frequency $\omega_{lmn}$ is unique to the spin and mass of a remnant black hole (for a review of black hole QN modes, see e.g. Ref. \cite{Berti:2009kk}). Soon after the merger of two progenitor black holes, the system is described by the Kerr solution with small perturbations. Indeed, it was proposed that the excitation of QN modes can be seen around the strain peak of GW signal for comparable mass mergers \cite{Giesler:2019uxc}. Then it was recognized \cite{Ioka:2007ak,Nakano:2007cj,Okuzumi:2008ej,Cheung:2022rbm,Mitman:2022qdl} that non-linearities, i.e., the second-order or even higher-order perturbations, are also important to precisely model merger-ringdown waveforms as was studied in Ref. \cite{Gleiser:1995gx,Gleiser:1996yc,Gleiser:1998rw,Campanelli:1998jv,Nicasio:2000ge,Brizuela:2006ne,Ioka:2007ak,Sberna:2021eui,Redondo-Yuste:2023ipg,Zhu:2024rej}.
Although ringdown and the excitation of QN modes are well understood, there is an unavoidable issue so-called {\it overfitting} problem (see e.g., Refs.~\cite{Baibhav:2023clw,Clarke:2024lwi}) if the ringdown starts around the strain peak and several QN modes, including the quadratic QN modes, should be taken into account in the model. It is caused by having many fitting parameters in the ringdown model.\footnote{One could extract the amplitude of each QN mode while avoiding the overfitting by making sure the stability of extracted amplitude against the change of the assuemd start time of ringdown \cite{Giesler:2019uxc,Baibhav:2023clw,Clarke:2024lwi}.} Also, the time-shift problem \cite{Sun:1988tz,Nollert:1998ys,Andersson:1996cm} is another main issue in the model of superposed QN modes as we have to guess the start time of ringdown to perform the data analysis for ringdown. To avoid these problems, one of the authors has recently proposed a model alternative to the superposed QN modes \cite{Oshita:2023cjz}, by which the spectral amplitude of ringdown can be modeled only with the reflectivity of the black hole light ring ${\cal R}_{lm}(\omega)$  or the black hole greybody factor, which is a function of mode frequency $\omega$. 

\begin{figure}[b]        \includegraphics[width=0.85\linewidth]{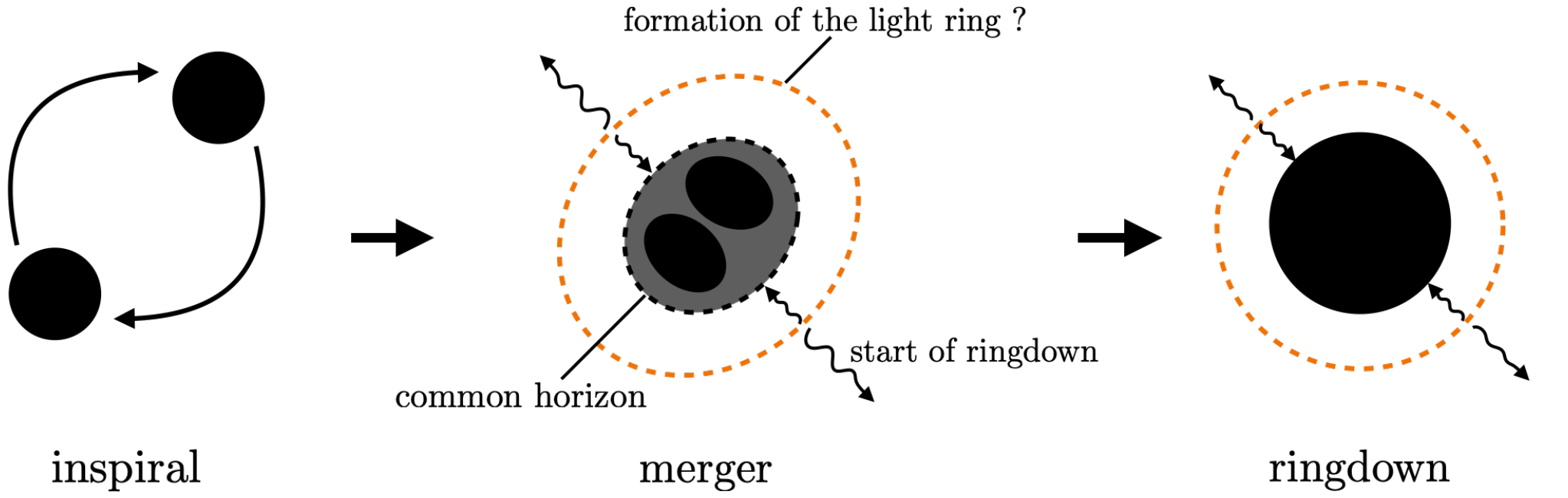}
        \caption{
        A schematic picture describing the conjectured scenario that the remnant light ring forms as early as in the merger phase of a BBH system.
        }
        \label{fig:schematic}
\end{figure}
The proposed greybody factor model for ringdown does not use individual QN-mode frequencies and does not include fitting parameters except for the overall amplitude. The reflectivity ${\cal R}_{lm}$ is represented by the greybody factor $\Gamma_{lm}$, i.e., transmissivity of the light ring, as ${\cal R}_{lm} = \sqrt{1-\Gamma_{lm}}$. This model has some limitations, e.g., it works when the source term has its small dependence on $\omega$ at $\omega \gtrsim f_{lm}\coloneqq f_{lm0}$, but the number of fitting parameters is significantly reduced. The frequency region relevant to ringdown, i.e., $\omega \gtrsim f_{lm} (M,a)$, depends only on the remnant parameters, mass $M$ and spin $a$. On the other hand, the data analysis of ringdown in the time domain involves an uncertainty in the start time of ringdown. In this sense, modeling of ringdown with the greybody factor has some advantages in the extraction of remnant parameters and in the test of the no-hair theorem of black holes without the uncertainty in the range of time-domain data and without overfitting that may affect data analysis. Also, the QNM filtering \cite{Ma:2022wpv,Ma:2023vvr} is recently proposed as a technique to erase the excitation of QN modes. It is an important technique as it also does not require fitting parameters and only requires properly adjusting the start time of ringdown. We should note that in the frequency domain, the amplitude of the merger-ringdown part includes a small amount of the inspiral GW, which can be contamination in the greybody factor model.

The ringdown model based on the greybody factor $\Gamma_{lm}$ was investigated only for extreme mass-ratio mergers \cite{Oshita:2023cjz}. In this paper, we investigate the feasibility of modeling ringdown with greybody factors for the case of comparable mass mergers in a phenomenological manner. 
To this end, we fit the reflectivity of a Kerr black hole ${\cal R}_{22}$ with the SXS's spectral amplitude of ringdown for the quadrupole moment. We will also fit the greybody factor model with the SXS waveforms with various mass ratios and remnant spin parameters to see the limitations of the model. 
We then demonstrate that the best-fit remnant parameters obtained with the greybody factor model are well consistent with the true remnant parameters.
We also propose another model in which the post-Newtonian (PN) correction is phenomenologically included and can model the GW spectral amplitude at frequencies slightly lower than $f_{22}$. The greybody factor model with the PN correction may increase the signal-to-noise ratio to improve the accuracy of the mass-spin extraction.
In the previous work \cite{Nichols:2010qi}, it was implied that GW waveform of the head-on collision of two black holes seen at a distant observer can be approximately captured by the PN correction and the black hole perturbations only. The study was also extended to the inspiralling case as well in Ref.~\cite{Nichols:2011ih}.

We here consider GW signals sourced by comparable mass-ratio binary black hole (BBH) mergers involving a highly non-linear phase. On the other hand, our phenomenological model implies that the spectral amplitude of the merger-ringdown phase can be modeled by the greybody factor based on the linear perturbation theory. For the consistency between the two views, we conjecture that the light ring forms at a very early stage of the merger (see Figure \ref{fig:schematic}). Relevant proposals have been made from different points of view, e.g., the Effeective One-Body method \cite{Buonanno:2000ef}, a method of matching post-Newtonian and black-hole-perturbation theories on a timelike surface \cite{Nichols:2010qi,Nichols:2011ih}, and the Backwards One-Body method \cite{McWilliams:2018ztb}. It is similar to the formation of the outermost common horizon in the merger phase \cite{Pook-Kolb:2019iao, Pook-Kolb:2019ssg, Mourier:2020mwa}. Although there are some relevant proposals in the definition of the light ring in more general cases \cite{claudel, siino_2019, siino_2021, siino2023black, yoshino_dtts, qasilocal_ps, PhysRevD.105.104040, amo2023generalization}, it is difficult to define the formation of the light ring in the dynamical and less-symmetric spacetime. To get insight into the formation of the light ring during a merger, we then consider a very simple solution, which is {\it static} but still less-symmetric one, the Majumdar-Papapetrou (MP) solution \cite{PhysRev.72.390, 10.2307/20488481}.
We discuss a scenario that the remnant light ring forms as early as the separation of the two equal-mass black holes is of the order of the horizon size.
Our discussion on the formation of the light ring in the simplest MP solution might shed light on how the separation of merging two black holes can affect the formation of the light ring. Note that a conclusive statement on the dynamical formation of the light ring requires a careful study of dynamical BBH solutions.

Our paper is organized as follows. In Sec. \ref{sec_review}, we review how we can model the spectral amplitude of GW signals sourced by BBH mergers.
We also explain how we obtain the spectral amplitude of SXS waveforms.
In Sec. \ref{sec_greybody_model}, we introduce the greybody factor model for our merger-ringdown phase based on the previous proposal \cite{Oshita:2023cjz} and explain the difference between the previous model \cite{Oshita:2023cjz} and ours.
In Sec. \ref{sec_exp_damp}, we carefully study how well our model matches with the SXS waveforms at higher frequencies relevant to merger-ringdown signals, i.e., $\omega \gtrsim f_{22}$.
In Sec. \ref{sec_mass_spin_measurement}, we compute the mismatch between a SXS waveform and our model for various assumed remnant parameters and show that the remnant parameters leading to the least value of the mismatch are consistent with the true remnant values.
In Sec. \ref{sec_PN_expansion}, we include the PN correction into the greybody factor model. We then demonstrate that it indeed works well to model the GW spectral amplitude not only higher but also lower frequencies as well only with a few fitting parameters. Also, the mass-spin extraction works slightly better than the greybody factor model without the PN correction.
In Sec. \ref{sec_MP_solution_interpretation}
, with a simple analytic solution, the MP solution, we discuss how the light ring of the remnant black hole forms and how 
the size of the light ring approaches that of the remnant one with respect to the separation of the binary.
In Sec. \ref{sec_conclusion}, we conclude and discuss the validity and limitations of our model and possible extensions of our work. We use the natural units $c=1$ and $G=1$ throughout the manuscript.

\section{Modeling ringdown for a BBH merger by greybody factors}
\label{sec_modeling_ringdown}

\subsection{spectral amplitude of GWs from a BBH: a brief review}
\label{sec_review}
The merger-ringdown GW waveforms sourced by comparable mass mergers can be obtained by numerical relativity. 
We here use SXS catalog \cite{Boyle:2019kee} to check our greybody factor model for merger-ringdown amplitudes. The SXS collaboration provides GW waveforms in normalized strain $r h_{lm}$ for each multipole moment. The strain data is decomposed into the real and imaginary parts $\text{Re/Im}(h_{lm}) \equiv h^{\rm (Re/Im)}_{lm}$ with various resolution levels. We here use SXS data extracted at $r/M_0=100$ with their highest available resolution, where the sum of the two Christodoulou masses $M_0$ is defined at $t=0$.

\begin{figure}[h]
\centering
\includegraphics[width=0.5\linewidth]{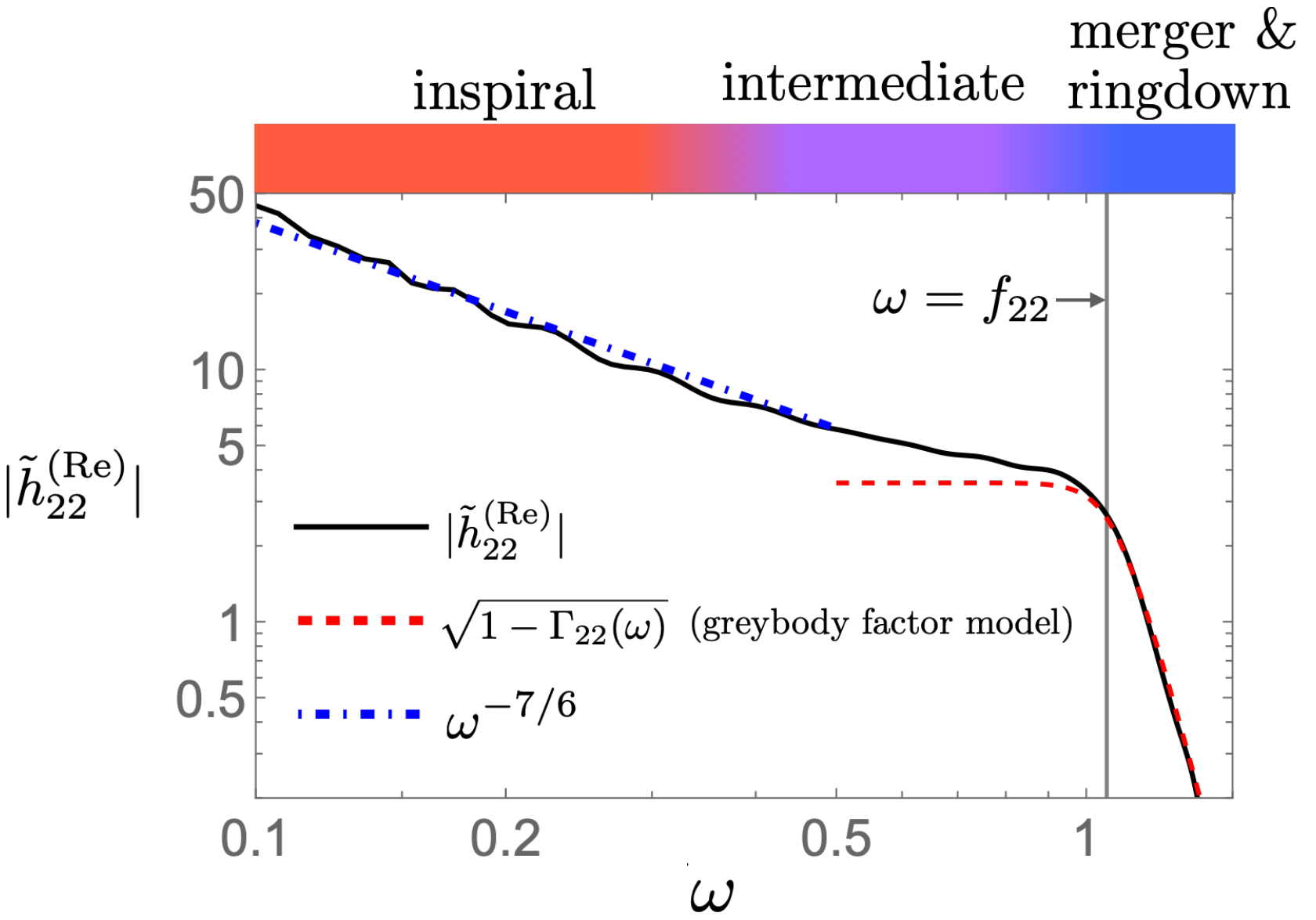}
\caption{GW spectral amplitude $|\tilde{h}_{22}^{\rm (Re)}|$ for SXS:BBH:0305. We set $t_i=100$ and $t_f=4457$ in (\ref{fourier_amplitude}). The amplitude of the inspiral phase can be approximated by $\omega^{-7/6}$ (blue dot-dashed) based on the stationary phase approximation. Merger and ringdown phases correspond to the high-frequency domain at $\omega \gtrsim f_{22}$. Our model based on the greybody factor $\Gamma_{22}$ is shown with the red dashed line.}
\label{fig:imr}
\end{figure}

A BBH merger, with the masses of the progenitors $M_1$ and $M_2$, emits a GW signal consisting of three phases: inspiral, merger, and ringdown (Figure \ref{fig:imr}). Note that Figure \ref{fig:imr} shows the spectral amplitude for the real part of the strain amplitude, i.e., $r h_{22}^{\rm (Re)}$, for SXS:BBH:0305. The spectral amplitude is computed by the following Fourier transform in the data range of $t_i \leq t \leq t_f$:
\begin{equation}
\tilde{h}_{22}^{\rm (Re/Im)} (\omega) = \int_{t_i}^{t_f} dt e^{-i\omega t} h_{22}^{\rm (Re/Im)} (t),
\label{fourier_amplitude}
\end{equation}
where we set $t_i =100$ and $t_f$ is set to the maximum time available in the SXS catalog. Also, we normalize the frequency in the unit of $2M=1$ throughout the manuscript. The values of a remnant mass $M$ and a remnant spin $j$ are available in the SXS catalog.
In the following, we often omit the superscript (Re/Im). The inspiral phase contributes to the low-frequency part of GW spectral amplitude. Based on the adiabatic analysis, an inspiral GW waveform is approximated as
\begin{equation}
h_{22} \sim \frac{\mu}{r} \left( \frac{1}{5} \frac{t_c -t}{\mu} \right)^{-1/4} \cos\left[{-2 \left( \frac{1}{5} \frac{t_c-t}{\mu} \right)^{5/8} + \delta}\right],
\end{equation}
where ${\mu} \equiv (M_1 M_2)^{3/5}/(M_1+M_2)^{1/5}$, $t_c$ is the time of coalescence and $\delta$ is a phase. The amplitude of this in the frequency domain is also analytically obtained by using the stationary phase approximation as
\begin{equation}
|\tilde{h}_{22}| \sim \frac{\mu^{5/6}}{r} \omega^{-7/6}.
\end{equation}
One can see that the spectral amplitude indeed follows $|\tilde{h}_{22}| \propto \omega ^{-7/6}$ at lower frequencies (blue dot-dashed line in Figure \ref{fig:imr}).
On the other hand, the contribution of the merger and ringdown phase comes at higher frequencies, where there are two characteristic features: a cut-off frequency $\omega \simeq f_{22} \coloneqq \text{Re}(\omega_{220})$ (grey solid line in Figure \ref{fig:imr}) and an exponential damping of spectral amplitude at $\omega \gtrsim f_{22}$. Our main argument is that those two features in the quadrupole spectral amplitude for BBH mergers can be modeled by the greybody factor $\Gamma_{22}$ at $\omega \gtrsim f_{22}$, i.e., merger and ringdown part in the GW spectrum (red dashed line in Figure \ref{fig:imr}). In the next section, we will briefly review the greybody factor $\Gamma_{lm}$ and how the greybody factor model works for merger-ringdown waveforms sourced by comparable mass-ratio BBH mergers.
\subsection{a ringdown model based on greybody factors}
\label{sec_greybody_model}
\begin{figure}[t]
\centering
\includegraphics[width=0.95\linewidth]{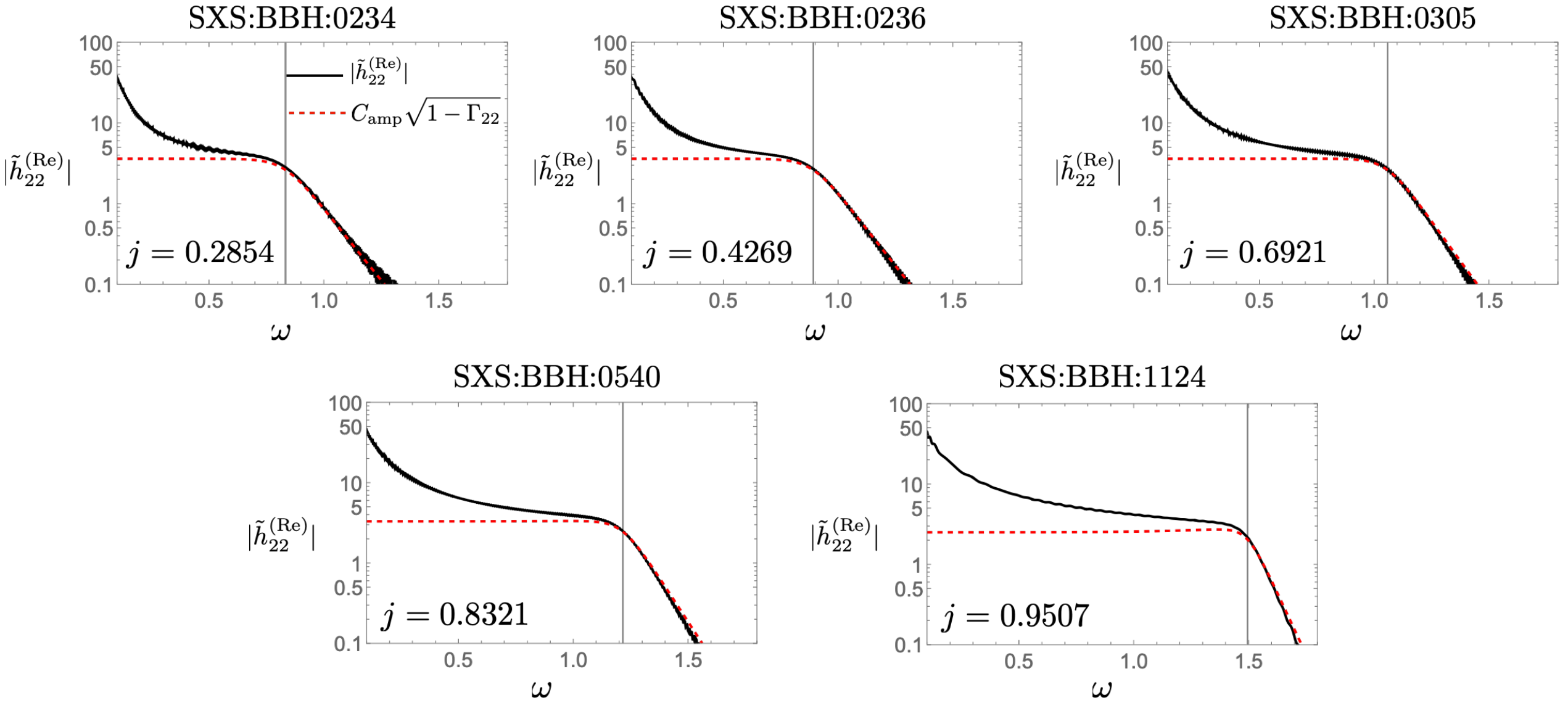}
\caption{Comparison between the greybody factor model (red dashed) and the spectral amplitude of GW from the SXS catalog (black solid). The vertical thin lines indicate the value of $f_{22}$. Mass ratio is in the range of $1\leq M_1 / M_2 \leq 2$.}
\label{fig:greybody_multicomp}
\end{figure}

The black hole greybody factor is one of the important quantities characterizing black holes. Given the mass, spin and charge of a black hole, the greybody factor $\Gamma_{lm} (\omega)$ is uniquely determined like QN modes.
It quantifies the transmissivity of the black hole's geometry for a mode frequency $\omega$ with a multipole moment $(l,m)$. As the greybody factor is obtained from the analysis of the scattering amplitude in the black hole's geometry, let us consider the Sasaki-Nakamura (SN) equation:
\begin{equation}
\left[ \frac{d^2}{dr^{\ast}{}^2} -F_{lm} \frac{d}{dr^{\ast}} -U_{lm} \right] X_{lm} = T_{lm},
\end{equation}
where $F_{lm}$ and $U_{lm}$ are given in the original paper \cite{Sasaki:1981sx} and $T_{lm}$ is the source term in the SN formalism.
We here introduced a new variable $r^{\ast}$ which is defined by
\begin{equation}
r^{\ast} \equiv r + \frac{1}{r_+ - r_-} \left[ r_+ \ln \left( r-r_+ \right) - r_- \ln \left( r-r_- \right) \right].
\end{equation}
Note that $r^{\ast} \to -\infty$ and $r^{\ast} \to + \infty$ corresponds to the horizon limit $r \to r_+$ and the far limit $r \to \infty$, respectively.
As the asymptotic behaviour of $F_{lm}$ and $U_{lm}$ in the SN equation is
\begin{equation}
F_{lm} \to 0 \ \text{for} \ r^{\ast} \to \pm \infty,
\end{equation}
and
\begin{align}
U_{lm} \to
\begin{cases}
-\omega^2 \ &\text{for} \ r^{\ast} \to \infty,\\
-k_{\rm H}^2 \ &\text{for} \ r^{\ast} \to -\infty,
\end{cases}
\end{align}
the asymptotic form of the homogeneous solution to the SN equation $X_{lm}^{\rm (hom)}$ is given by the linear combination of the two independent solutions, $X_{lm}^{\rm (in)}$ and $X_{lm}^{\rm (out)}$:
\begin{align}
X_{lm}^{\rm (in)} =
\begin{cases}
A_{\rm in}^{(l,m)} (\omega) e^{-i\omega r^{\ast}} + A_{\rm out}^{(l,m)} (\omega) e^{i\omega r^{\ast}} &\text{for} \ r^{\ast} \to \infty,\\
e^{-ik_{\rm H} r^{\ast}} &\text{for} \ r^{\ast} \to -\infty,
\end{cases}
\end{align}
and
\begin{align}
X_{lm}^{\rm (out)} =
\begin{cases}
e^{i\omega r^{\ast}} &\text{for} \ r^{\ast} \to \infty,\\
B_{\rm out}^{(l,m)} (\omega) e^{ik_{\rm H} r^{\ast}} + B_{\rm in}^{(l,m)} (\omega) e^{-ik_{\rm H} r^{\ast}} &\text{for} \ r^{\ast} \to -\infty,
\end{cases}
\end{align}
where $k_{\rm H} \equiv \omega - m \Omega_{\rm H}$, $\Omega_{\rm H} \equiv j / 2 r_+$ and $A_{\rm in/out}^{(l,m)}$ and $B_{\rm in/out}^{(l,m)}$ are coefficients depending on $\omega$. The greybody factor is identical to the absorption probability of an incoming mode $\omega$. Given the coefficients $A_{\rm in}^{(l,m)}$ and $A_{\rm out}^{(l,m)}$, the greybody factor is given by
\begin{equation}
\Gamma_{lm} = \left|\frac{C}{c_0}\right|^2 \left|\frac{A_{\rm out}}{A_{\rm in}} \right|^2, 
\end{equation}
where $|C|^2$ and $c_0$ \cite{Starobinsky:1973aij,Starobinsky_Churilov,Teukolsky:1974yv} are
\begin{align}
|C|^2 &\equiv \lambda^4 + 4 \lambda^3 + \lambda^2 (-40 a^2 \omega^2 + 40 am\omega + 4) + 48 a \lambda \omega (a \omega + m) + 144 \omega^2 (a^4 \omega^2 -2 a^3 m \omega + a^2 m^2 + 1/4),\\
c_0 &\equiv \lambda (\lambda + 2) -12 a \omega (a \omega - m) -6i \omega,
\end{align}
where $\lambda$ is the separation constant leading to the regular spheroidal harmonics.
In this paper, as shown in Figure \ref{fig:imr}, we introduce a phenomenological model for the merger-ringdown spectral amplitudes with the quadrupole moment:
\begin{equation}
|\tilde{h}_{22}^{\rm (Re/Im)} (\omega)| = C_{\rm amp} \sqrt{1-\Gamma_{22} (\omega)}, \ \text{for} \ \omega \gtrsim f_{22},
\end{equation}
where a constant $C_{\rm amp}$ is an overall amplitude. For comparable mass mergers, the greybody factor model indeed works well for various spin parameters as is shown in Figure \ref{fig:greybody_multicomp}.

In the following sections, we will demonstrate that the spectral amplitude of the merger and ringdown phase for $(l,m)=(2,2)$ can be modeled by a universal quantity, $\Gamma_{22} (\omega)$. We also demonstrate that our greybody factor model works to infer the remnant parameters.

\subsection{exponential damping of the merger-ringdown spectral amplitude}
\label{sec_exp_damp}
We here study how well our model fits to the SXS data at higher frequencies where the spectral amplitude is exponentially damped with $e^{-(\omega - f_{22})/(2T^{\rm (GW)})}$. A constant $1/T^{\rm (GW)}$ is the strength of the damping in the spectral amplitude. Our argument is that the damping amplitude can be modeled by $C_{\rm amp} \sqrt{1-\Gamma_{22}(\omega)}$, which has a damping $\sim e^{-(\omega - f_{22})/(2T)}$ at $\omega \gtrsim f_{22}$. Here $C_{\rm amp}$ is an overall amplitude. We will see how $T^{\rm (GW)}$ extracted from SXS waveforms is consistent with the value of $T$ in the greybody factor.\footnote{A similar analysis for extreme mass-ratio mergers has been performed in Ref. \cite{Oshita:2022pkc} and \cite{Oshita:2023cjz} by one of the authors.}
\begin{figure}[t]
\centering
\includegraphics[width=0.95\linewidth]{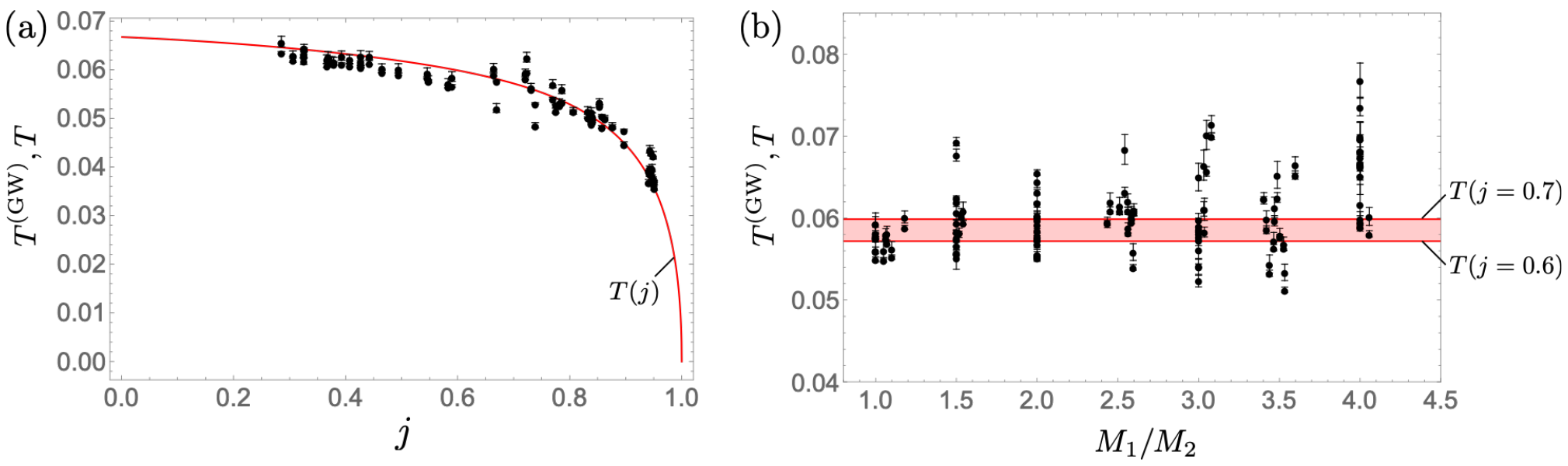}
\caption{(a) Exponent parameter $T^{\rm (GW)}$ of the damping amplitude $|\tilde{h}_{22}| \sim e^{-\omega/(2 T^{\rm (GW)})}$ at $\omega \geq \alpha f_{22}$ is shown with respect to the remnant spin $j$ (black markers). The constant $\alpha$ determines the cut-off frequency, and how we determine $\alpha$ is described in Appendix \ref{app_extraction}. The mass ratio is in the range of $1 \leq M_1/M_2 \leq 2$. The red solid line indicates the exponent of the damping $T$ in $\sqrt{1-\Gamma_{22}} \propto e^{-\omega/(2T)}$ in $\omega \gtrsim f_{22}$. (b) Exponent parameter $T^{\rm (GW)}$ are shown with respect to the mass ratio $M_1/M_2$ (black markers). The mass ratio is in the range of $1 \leq M_1/M_2\lesssim 4$ and the spin parameter is in the range of $0.6 \leq j \leq 0.7$. The red band indicates the range of $T(j=0.7)\leq T \leq T (j=0.6)$.
In (a) and (b), we use 55 and 70 SXS waveforms, respectively, and extract $T^{\rm (GW)}$ independently from each polarization $h_{+}$ and $h_{\times}$.}
\label{fig:spin_comp}
\end{figure}
The fitting function for the exponent parameter $T=T(j)$ with $2M=1$ is given in Ref. \cite{Oshita:2023cjz}. We show the comparison between $T$ and $T^{\rm (GW)}$ in Figure \ref{fig:spin_comp}.
Figure \ref{fig:spin_comp}-(a) shows the values of $T$ and $T_{\rm GW}$ with respect to the remnant spin, and the mass ratio is restricted to a comparable mass ratio of $1 \leq M_1/M_2 \leq 2$. On the other hand, in Figure \ref{fig:spin_comp}-(b), the comparison is performed with respect to the mass ratio of $1 \leq M_1/M_2 \lesssim 4$, and the remnant spin is restricted to the typical values of $0.6 \leq j \leq 0.7$.
We normalize the spectral amplitude of SXS data with the scale of $2M=1$ and extract the exponent parameter $T^{\rm (GW)}$ by using the \texttt{NonlinearModelFit} in the Mathematica. The error is estimated by Mathematica's option \texttt{ParameterErrors} which gives the standard errors for parameter estimates.

From Figure \ref{fig:spin_comp}-(a), we find that the exponent parameter $T^{\rm (GW)}$ extracted from the SXS data is consistent with the exponent of the damping in $\sqrt{1-\Gamma_{22} (\omega)}$ in the wide range of the spin parameter. From Figure \ref{fig:spin_comp}-(b), on the other hand, we find that the exponent of the damping in the reflectivity agrees with $T^{\rm (GW)}$, especially for comparable mass ratios $M_1/M_2 \simeq 1$. For larger mass ratios, the exponent of $T^{\rm (GW)}$ tends to be larger than $T$. This may be caused by the sourcing effect of a small companion black hole. Indeed, in the previous work done by one of the authors \cite{Oshita:2023cjz}, one needs an extra factor of $1/\omega^3$ in the model for the ringdown spectral amplitude for extreme mass ratios.

Our model could be applied to another model for the merger-ringdown amplitude proposed in Refs. \cite{Khan:2015jqa,Husa:2015iqa}. They model the exponential damping in the amplitude with $\tau_{lm0} \times \gamma_3$ where $\gamma_3$ is a free parameter of order unity \cite{Khan:2015jqa,Husa:2015iqa}. We here do not use extra free parameters except for the overall amplitude $C_{\rm amp}$.
\subsection{mass-spin measurement}
\label{sec_mass_spin_measurement}
\begin{figure}[t]
\centering
\includegraphics[width=0.8\linewidth]{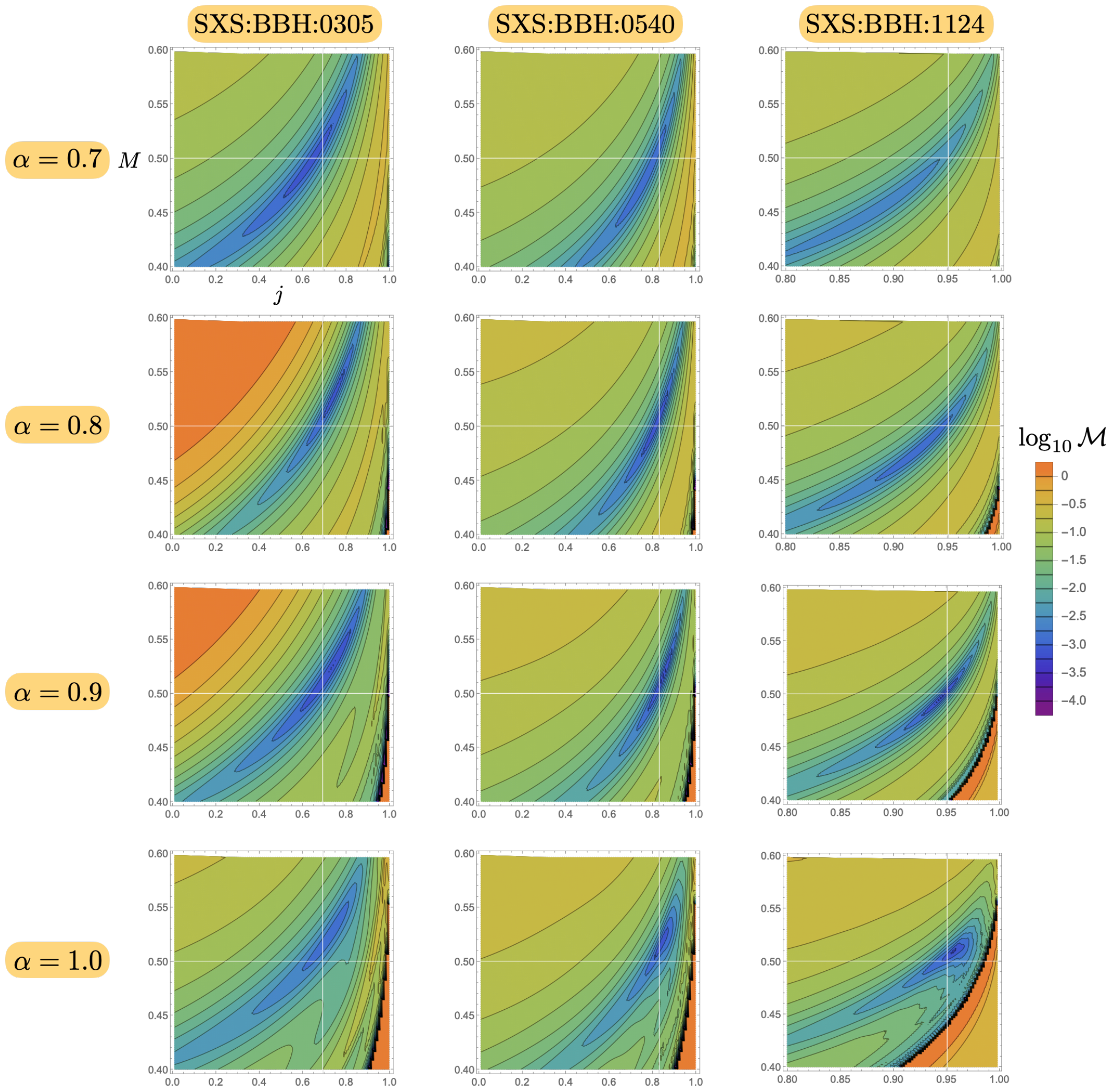}
\caption{Mismatch between our model $C_{\rm amp}{\cal R}$ and the spectral amplitude of the SXS's waveforms, SXS:BBH:0305 ($j=0.6921$), SXS:BBH:0540 ($j=0.8321$), and SXS:BBH:1124 ($j=0.9507$). The white lines indicate the true value of the remnant quantities and the remnant mass $M_{\rm true}$ is normalized with $M_{\rm true}=0.5$. We set $\omega_f = 1.7$ and $\alpha = 0.7$, $0.8$, $0.9$, and $1.0$.}
\label{fig:alpha_change}
\end{figure}

In this section, we show how our phenomenological model for the merger-ringdown amplitude works to infer the remnant mass and spin from SXS's waveforms.
\begin{figure}[t]
\centering
\includegraphics[width=1\linewidth]{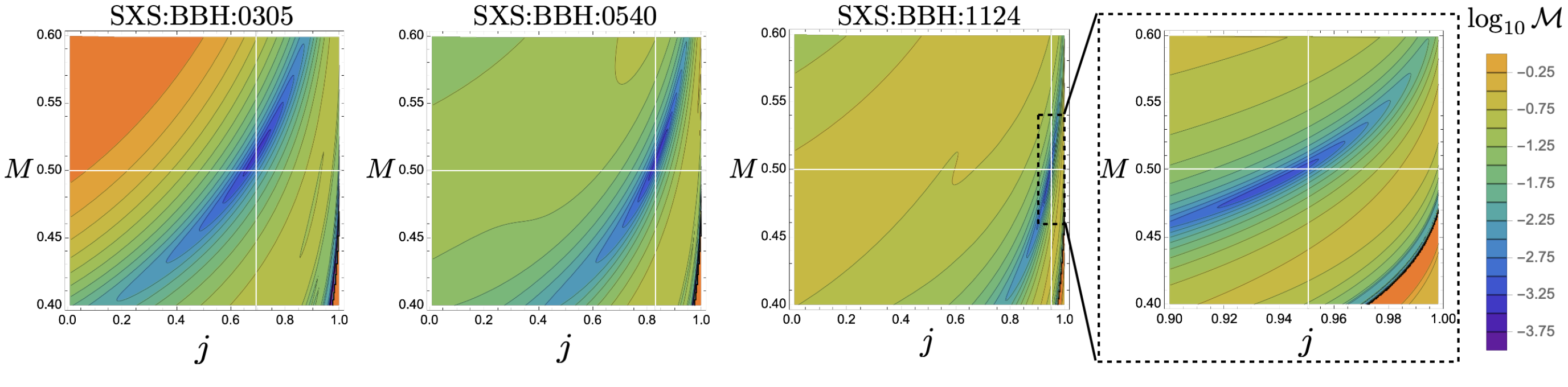}
\caption{Mismatch between our model $C_{\rm amp}{\cal R}$ and the spectral amplitude of the SXS's waveforms, SXS:BBH:0305 ($j=0.6921$), SXS:BBH:0540 ($j=0.8321$), and SXS:BBH:1124 ($j=0.9507$). The white lines indicate the true value of the remnant quantities and the remnant mass $M_{\rm true}$ is normalized with $M_{\rm true}=0.5$. We set $\omega_i = 0.85 \times f_{22}(M,a)$ and $\omega_f = 1.7$ for all the three SXS waveforms.}
\label{fig:mismatch}
\end{figure}
To this end, we compute mismatch ${\cal M}$ between the SXS data, $|\tilde{h} (\omega)|$, and our model for the merger-ringdown amplitude $C_{\rm amp} {\cal R} (M,j) = C_{\rm amp} \sqrt{1-\Gamma(M,j,\omega)}$. We then find the best-fit remnant parameters ($M,j$) for which the mismatch takes the least value. We estimate the mismatch with the following formula
\begin{equation}
{\cal M} (M,j) = \left|1- \frac{\braket{|\tilde{h}||C_{\rm amp} \times {\cal R}}}{\sqrt{\braket{|\tilde{h}|||\tilde{h}|}\braket{C_{\rm amp} \times  {\cal R}|C_{\rm amp} \times {\cal R}}}} \right| = \left|1- \frac{\braket{|\tilde{h}||{\cal R}}}{\sqrt{\braket{|\tilde{h}|||\tilde{h}|}\braket{{\cal R}|{\cal R}}}} \right|,
\end{equation}
where $\braket{a(\omega) | b(\omega)}$ is
\begin{equation}
\braket{a(\omega) | b(\omega)} = \int_{\omega_i}^{\omega_f} d\omega a(\omega) b^{\ast}(\omega),
\label{inner_prodct}
\end{equation}
and $\omega_i$ and $\omega_f$ determine the data range in the frequency domain used to compute the mismatch. Our results are shown in Figure \ref{fig:mismatch}.
We here take the spectral data at $\omega \geq \omega_{i} = \alpha \times f_{22}(M, a)$ with $\alpha=0.85$ throughout the mass-spin inference we performed for the three SXS waveforms, i.e., we do not tune the value of $\alpha$ for the different waveform. Nevertheless, we see that the true remnant values (white lines in Figure \ref{fig:mismatch}) are consistent with the best-fit parameters where ${\cal M}$ takes the least value.
We find that the value of $\alpha \simeq 0.85$ works for other SXS waveforms (see Appendix \ref{app:mass_spin_noPN}), although the feasibility of the extraction of the remnant mass and spin is not very sensitive to the choice of $\alpha$ as can be seen in Figure \ref{fig:alpha_change}. The mass-spin parameter region where ${\cal M} \lesssim 10^{-3}$ is consistent with the true remnant values for $0.8 \lesssim \alpha \lesssim 1$.

\subsection{greybody factor model with the Post-Newtonian correction}
\label{sec_PN_expansion}
\begin{figure}[h]
\centering
\includegraphics[width=0.8\linewidth]{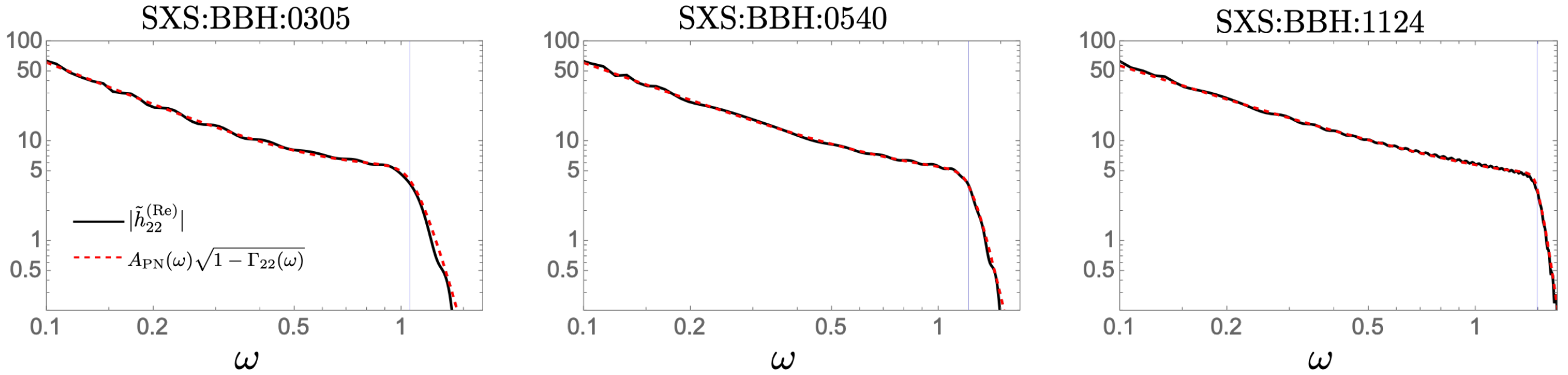}
\caption{The model of the greybody factor with the PN correction, $A_{\rm PN} \sqrt{1- \Gamma_{22}}$ is fitted to the spectral amplitude of SXS data. We here truncate $A_{\rm PN}$ at $k=4$, i.e., this has three fitting parameters, and perform the least-square fit. The blue vertical lines indicate $\omega = f_{22}$.}
\label{fig:fit_PN}
\end{figure}

We here take into account the pre-merger amplitude of GW spectrum that could be represented by the PN expansion. \footnote{In Ref.~\cite{Nichols:2010qi}, it was implied that GW waveform of the head-on collision of two black holes seen at a distant observer can be approximately captured by the PN correction and the black hole perturbations only. The study was extended to the inspiralling case as well in Ref.~\cite{Nichols:2011ih}.} We then propose the greybody factor model with the PN correction, which is applicable to GW amplitude not only at higher frequencies but also at lower frequencies. This model is more insensitive to the data range, controlled by the parameter $\alpha$, as will be demonstrated later. As shown in Figure \ref{fig:imr}, the PN amplitude $A_{\rm PN}$ obtained from the stationary phase expression, $A_{\rm PN} \propto \omega^{-7/6}$ is consistent with the inspiral phase at lower frequencies. The stationary phase expression can be re-expanded as \cite{Santamaria:2010yb}
\begin{equation}
A_{\rm PN} \sim C_{\rm amp} \omega^{-7/6} \left( 1 + \sum_{k=2}^{5} p_k \omega^{k/3} \right),
\end{equation}
where $p_k$ is an expansion constant. As the reflectivity $\sqrt{1-\Gamma_{\ell m}}$ is unity at lower frequencies, one can indeed fit the GW spectral $|\tilde{h}_{22}|$ with the model $A_{\rm PN} (\omega) \sqrt{1-\Gamma_{22} (\omega)}$ at the lower and higher frequencies (Figure \ref{fig:fit_PN}).
We truncate higher-order PN corrections ($k\geq 4$) as they have a positive power of $\omega$ and would not contribute to the model of GW spectrum at lower frequencies $\omega \lesssim f_{22}$.
Also, the inclusion of many fitting parameters in a model may lead to the overfitting issue.

\begin{figure}[t]
\centering
\includegraphics[width=0.8\linewidth]{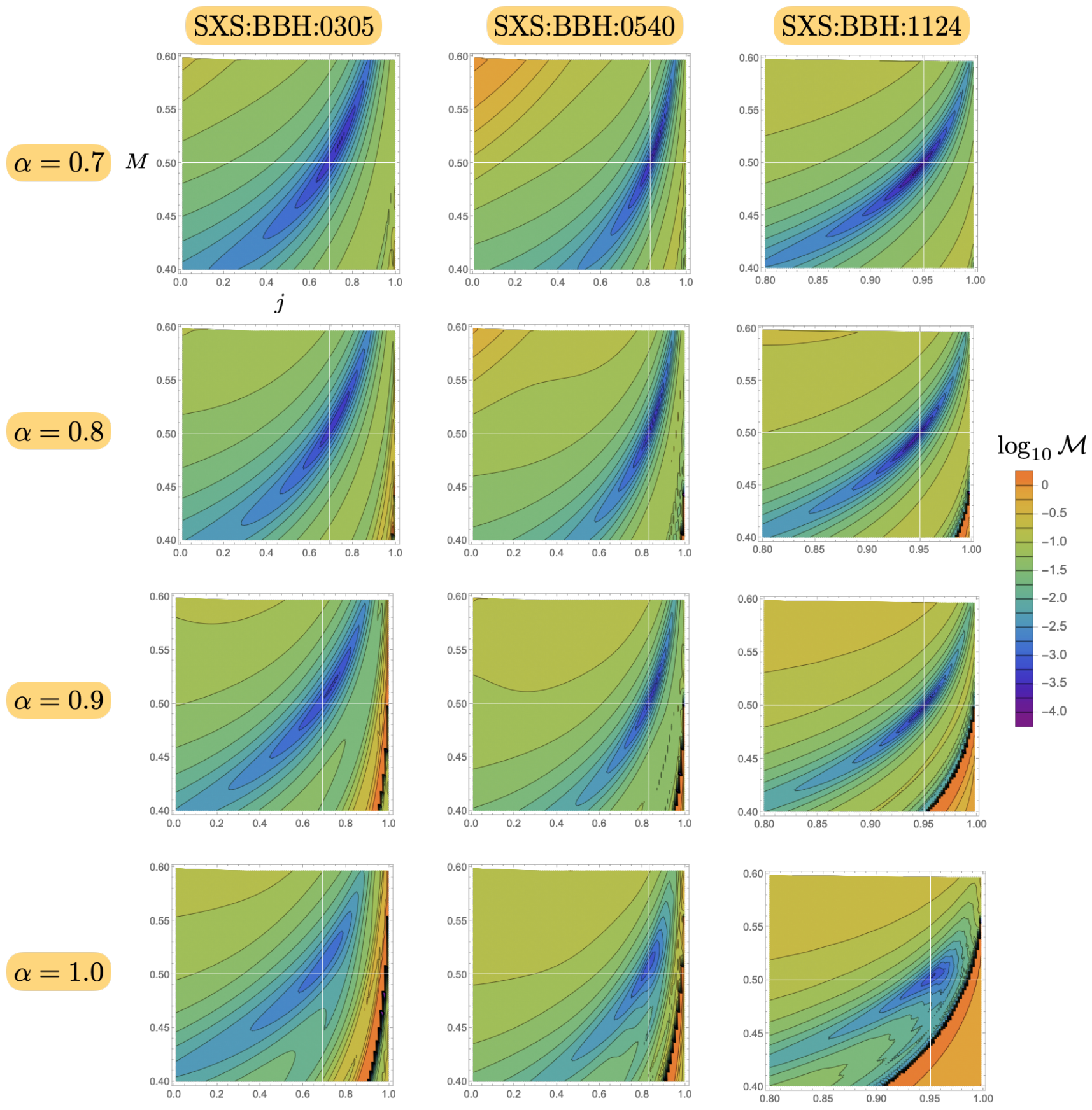}
\caption{Mismatch between the greybody factor model with the PN correction (\ref{greybody_PNexpansion}) and the spectral amplitude of the SXS's waveforms, SXS:BBH:0305 ($j=0.6921$), SXS:BBH:0540 ($j=0.8321$), and SXS:BBH:1124 ($j=0.9507$). The white lines indicate the true value of the remnant quantities and the remnant mass $M_{\rm true}$ is normalized with $M_{\rm true}=0.5$. We set $\omega_f = 1.7$ and $\alpha = 0.7$, $0.8$, $0.9$, and $1.0$.}
\label{fig:mass_spin_alpha_change_PN}
\end{figure}

To model the pre-merger amplitude of the spectral amplitude $|h_{\ell m}(\omega)|$ in the range of $0.5 f_{22} \leq \omega$ and to extract the remnant parameters from SXS waveforms, we consider the following model 
\begin{equation}
|\tilde{h}_{22}| \simeq (p_2 \omega^{-1/2} + p_3 \omega^{-1/6}) \sqrt{1 - \Gamma_{22}},
\label{greybody_PNexpansion}
\end{equation}
where we omit the term of $\omega^{-7/6}$ as it may decay and is less significant compared to the other PN corrections at the intermediate frequencies relevant to the merger and ringdown phase. 
In Figure \ref{fig:mass_spin_alpha_change_PN}, we compute the mismatch between the SXS data and the greybody factor model with the PN correction (\ref{greybody_PNexpansion}). To fix the parameters, $p_2$ and $p_3$, we perform the least-square fit in the frequency range of $0.5 f_{22} \leq \omega \leq 2$.
We find that the greybody factor model with the PN correction works well in the mass-spin measurement, compared with the previous greybody model without the PN correction (see Figures \ref{fig:alpha_change} and \ref{fig:mass_spin_alpha_change_PN}). Especially, the least value of the mismatch between the greybody factor model without the PN correction and SXS:BBH:1124 with $\alpha =0.7$ is off from the true value (Figure \ref{fig:alpha_change}), but the model with the PN correction still works in the same setup. It means that the inclusion of the PN correction makes our model more insensitive to the range of spectral data.\footnote{One may wonder how the PN correction affects the value of $T$. As the PN correction is the power of $\omega$, its effect can be neglected at high-frequency region ($\omega \gtrsim f_{22}$) where the exponential damp of the greybody factor dominates the $\omega$-dependence of the GW spectral amplitude.} For example, the mass-spin extraction from SXS data with the greybody factor model, including the PN correction, is stable within $0.7 \lesssim \alpha \lesssim 1.0$.\footnote{At the strain peak, several overtones may be highly excited with larger amplitudes as was demonstrated with the QN-mode fitting (e.g. in Ref. \cite{Giesler:2019uxc}) or as was shown by the excitation factors \cite{Oshita:2021iyn}. As the overtones have real-part frequencies smaller than that of the fundamental mode $f_{22}$, the excitation of overtones would be relevant not only to high-frequency spectrum ($\omega \gtrsim f_{22}$) but also to lower-frequency one ($\omega \sim \alpha f_{22}$).}
Also, it is confirmed that one can extract the consistent mass-spin values from other SXS waveforms with our greybody factor model with the PN correction as is shown in Appendix \ref{app:mass_spin_noPN}.

In summary, we list some advantageous/disadvantageous points of the greybody factor model for ringdown:
\begin{itemize}
\item[({\bf pros I})] The mass-spin measurement is stable against the change of data range (i.e. changing the $\alpha$) as is shown in Figures \ref{fig:alpha_change} and \ref{fig:mass_spin_alpha_change_PN} for without and with the PN correction, respectively. One also does not need to tune the data range for each GW waveform as we set the same value $\alpha=0.85$ for different SXS waveforms (see Figures \ref{fig:mismatch}, \ref{fig:085_noPN}, and \ref{fig:085_PN}). 
\item[({\bf pros II})] In the greybody factor model with the PN correction, there is only two fitting parameters, i.e., $p_2$ and $p_3$, except for the remnant spin and mass. The small number of fitting parameters in a model may enable us to avoid the overfitting problem in the QNM fitting analysis. Actually, the greybody factor includes the contribution of all overtones weighted with the excitation factors. This significantly reduces the number of fitting parameters at the cost of cons I mentioned below. 

\item[({\bf cons I})] The frequency dependence of the GW spectrum is determined not only by the greybody factor but also by the source term. The ambiguity in the source term can lead to systematic error, which could prevent us from extracting the correct greybody factor from GW data.
One can partially resolve this issue by including the pre-merger part that can be modeled with the PN correction (see Figure \ref{fig:fit_PN}).
The inclusion of the tail signals is difficult as it is associated with the branch point of the Green's function at $\omega=0$. The inclusion of quadratic QN modes or other non-linear effects will be the future work.

\item[({\bf cons II})] As the greybody factor has no information about the phase of the GW spectrum, this model cannot maximally utilize all available information in the GW data.
\end{itemize}

\section{Discussion: formation of the light ring in the static BBH}
\label{sec_MP_solution_interpretation}
One may wonder why the greybody factor, obtained in the linear perturbation regime, works to model the merger-ringdown waveform that may include more or less nonlinear effects. As a possible scenario, we would argue that even if the interior of the outermost common horizon is highly non-linear, its exterior instantaneously relaxes to the Kerr geometry with perturbations, and the remnant light ring forms there (see Figure \ref{fig:schematic}). This conjecture may be relevant to the previous work in Refs. \cite{Nichols:2010qi,Nichols:2011ih} where the authors provide some supporting evidence by using both the post-Newtonian and black hole perturbation techniques to describe GW emission sourced by a BBH merger involving nonlinear collision.
The relaxation of a merging BBH is complicated as it is less symmetric and highly dynamical.
There is no unique consensus on the definition of the light ring in the dynamical spacetime although it has been discussed how one can define it \cite{claudel, siino_2019, siino_2021, siino2023black, yoshino_dtts, qasilocal_ps, PhysRevD.105.104040, amo2023generalization}.

We here discuss the early formation of the light ring, which may be relevant to the greybody factor imprinted on the ringdown spectrum. Note that the main purpose of this paper is to apply the greybody factor model \cite{Oshita:2023cjz} to the SXS waveforms and to present our interpretation of why our model works to model the ringdown spectral amplitude even for binary black hole mergers as shown in Sec. \ref{sec_modeling_ringdown}. 

We here consider a binary system that is less symmetric but is static, i.e., has no dynamics. Such a system can be described by the Majumdar-Papapetrou solution \cite{PhysRev.72.390, 10.2307/20488481}.
For the case of less-symmetric and dynamical collisions, we need both numerical relativity and a reasonable definition of the light ring applicable even to dynamical systems, which is beyond the scope of our purpose of this paper. We are interested in how the separation of two merging black holes may affect the formation of the light ring. More concretely, we here consider the MP solution with the two extreme black holes of mass $M/2$ and control the separation of the binary denoted by $d$. When $d=0$, it leads to the remnant black hole of mass $M$ which has spherical symmetry, and the light ring is recovered. When $d \gg M$, it has no light ring enclosing the binary. At which separation does the remnant light ring form? How does the formed light ring approach the spherical form with respect to the separation $d$? The former question is relevant to the formation during the merger phase and the latter one would be relevant to the relaxation process of the light ring. It would be useful to discuss the formation of the light ring from this point of view although most of the facts relevant to the MP solution introduced here are known \cite{doi:10.1098/rspa.1989.0010, Shipley_2016, PhysRevD.98.064036, PhysRevD.95.024026}. However, we should note that the two factors, the separation and dynamics of a BBH, may not be independent of each other in a realistic binary system.

The MP spacetime is a solution of the Einstein-Maxwell system, and its line element $ds^2$ and the electromagnetic potential $A=A_{\mu} dx^\mu$ are given by
\begin{align}\label{eq:MPmetric}
    ds^2 &= -U^{-2}dt^2+ U^2(dx^2+dy^2+dz^2),\\
    A &= U^{-1} dt,
\end{align}
where
\begin{align}
    U &\coloneqq 1 + \sum_{i=1}^{N} \frac{M_i}{r_i},\\
     r_{i} &\coloneqq \sqrt{(x-x_i)^2+(y-y_i)^2+(z-z_i)^2}.
\end{align}
The metric describes a system of $N$ extremely charged ``particles'', and the $i$-th particle is located at $(x,y,z)=(x_i,y_i,z_i)$. Each mass and charge is denoted by $M_i$ which is a positive real value.

In the case of $N=2$ that describes a static BBH, the metric describes two extremely charged particles, and the spacetime is static due to the balance of the gravitational force and the electric force for each particle. We adopt the cylindrical coordinates $(\rho, \phi, z)$, in which the extremely charged particles are located at $z = \pm d$ on the $z$-axis. If the particles have the identical mass $M_1=M_2=M/2$, then the metric is written as 
\begin{equation}\label{eq:MPmetric_two_particles}
    ds^2 = -U^{-2}dt^2+ U^2(d\rho^2+\rho^2 d\phi^2 + dz^2),
\end{equation}
where
\begin{equation}
    U = 1 + \frac{ M }{ 2\sqrt{\rho^2 + (z-d)^2} }+ \frac{M}{2\sqrt{ \rho^2 + (z+d)^2 }}.
\end{equation}
The spacetime has the apparent coordinate singularities at $z = \pm d$, and the analytical continuation is possible at each point \cite{HH1972}. 
Actually, the points at $z = \pm d$ are the event horizons, and the horizon area at each point is the same as the extreme Reissner-Nordstr\"{o}m black hole with mass $M/2$. 
\begin{figure}[t]
        \includegraphics[width=0.9\linewidth]{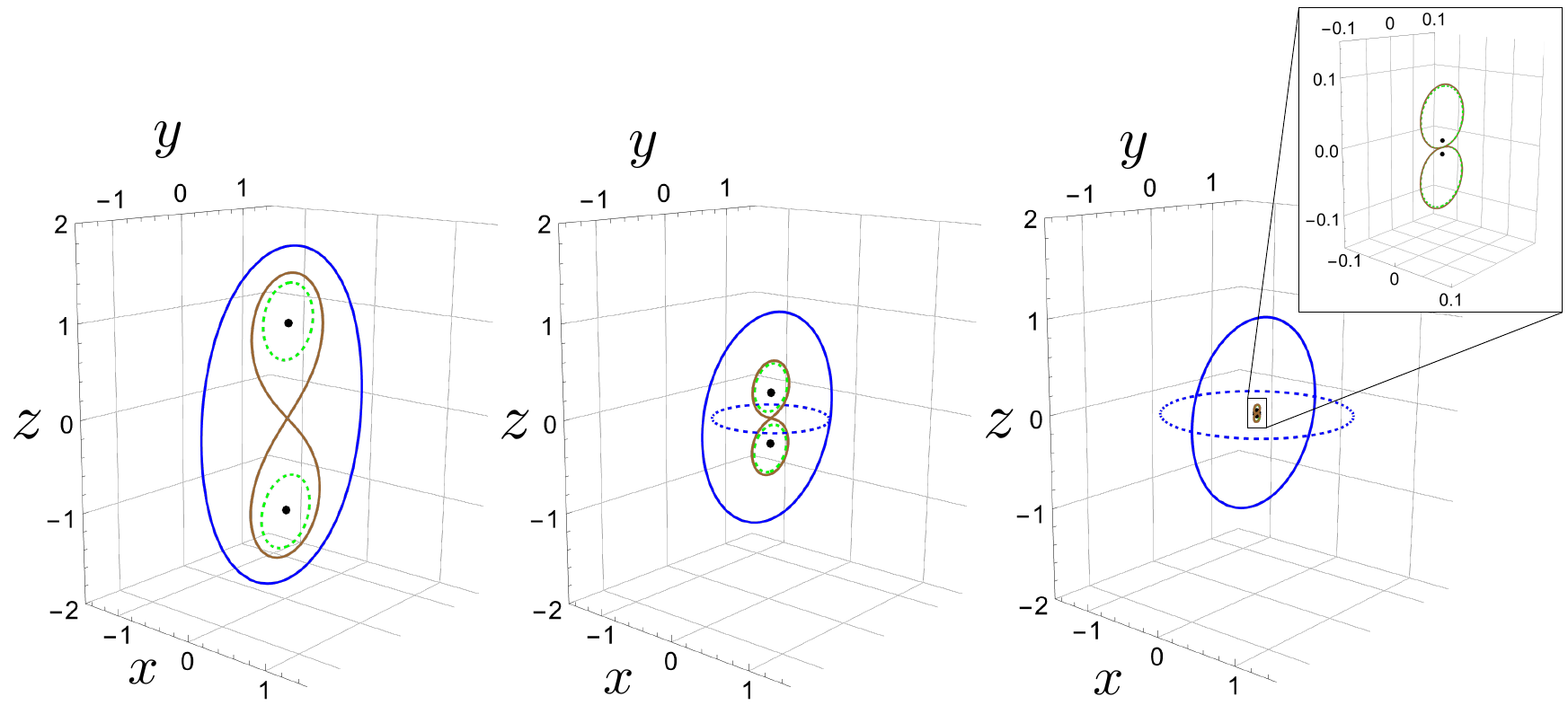}
        \caption{
        Three-dimensional plots of closed orbits on the equatorial and meridian plane in Majumdar-Papapetrou spacetime. We set the separation as $d=M$ (left), $d=d_{\rm crit}$ (center), and $d=10^{-2}M$ (right).
        On the meridian plane $(x=0)$, there are three types of closed orbits: an orbit enclosing two black holes (blue solid), an 8-shaped orbit (brown solid), and two orbits enclosing each black hole (green dashed). In the limit $d \rightarrow 0$, the blue orbit converges to the photon sphere of the extreme Reissner-Nordstr\"{o}m black hole while the other orbits on the meridian plane shrink to the origin.
        The top-right panel shows an enlarged view of the right panel.
        On the equatorial plane $(z=0)$, the dashed blue line depicts an unstable circular orbit. Since an unstable circular orbit on the equatorial plane does not exist for $d > d_{\rm crit}$, there is no circular orbit in the left figure ($d=M$). The dashed blue line on the equatorial plane converges to the photon sphere of the extreme Reissner-Nordstr\"{o}m black hole.
        }
        \label{fig:plot3D_closed_orbits}
\end{figure}
\begin{figure}[h]
\centering
\includegraphics[width=0.45\linewidth]{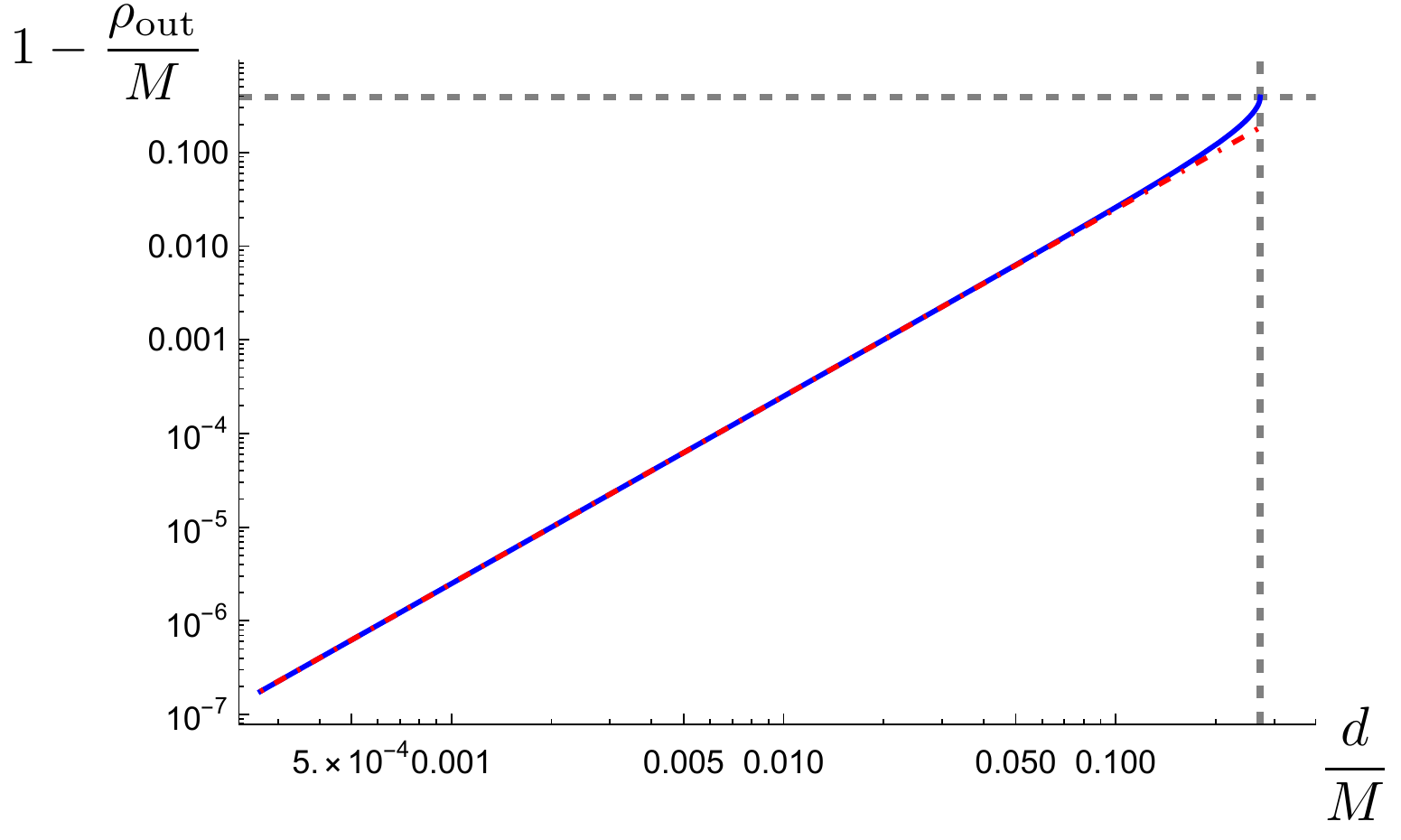}
\caption{The difference between the size of the outer photon sphere on the equatorial plane for $d > 0$ and that for $d=0$ (i.e. Reissnor-Nordstr\"{o}m black hole spacetime), quantified by $1-\rho_{\rm out}/M$, is shown with respect to the separation $d/M$. The blue solid line shows the analytical solution given by Eq. (\ref{eq:rho_out}) and the red dot-dashed line shows $(5/2)d^2/M^2$ (see Eq. (\ref{eq:expanded_rho_out})). The deviation converges to zero with $\propto d^2/M^2$.
}
\label{fig:radius_of_outer_circular_orbit}
\end{figure}
In the cylindrical coordinates of Eq.(\ref{eq:MPmetric_two_particles}), we consider the two types of closed null geodesics: (i) one on the equatorial plane ($z=0$) and (ii) another one is on the meridian plane ($\phi={\rm const.}$).
On the equatorial plane, there are two circular orbits if half of the coordinate distance between black holes $d$ is smaller than $d_{\rm crit}=\frac{M}{3}\sqrt{\frac{2}{3}}$. The outer orbit approaches the circular orbit in the extreme Reissnor-Nordstr\"{o}m black hole spacetime, whose radius is $M$, as $d \rightarrow 0$ while the inner orbit shrinks to the origin as $d \rightarrow 0$.
On the meridian plane, there are three types of closed orbits \cite{doi:10.1098/rspa.1989.0010, Shipley_2016, PhysRevD.98.064036}. One is the orbit enclosing the two black holes, the second one is the 8-shaped orbit, and the third ones are the orbits enclosing each black hole. The orbit enclosing the two black holes approaches the circular orbit in the extreme Reissnor-Nordstr\"{o}m black hole spacetime as $d \rightarrow 0$ while the others shrink to the origin as $d \rightarrow 0$. 
These closed null orbits on the equatorial plane and the meridian plane in terms of the coordinate distance $d$ are shown in Figure \ref{fig:plot3D_closed_orbits}. 
In the following, we analyze how the closed null orbits on the equatorial plane and the meridian plane approach the circular orbit in the extreme Reissnor-Nordstr\"{o}m black hole spacetime in terms of the coordinate distance $d$ instead of the proper distance. This is because the proper distance between the two black holes on the same time slice diverges due to their extreme charge.

On the equatorial plane, the geodesic equation is separable, and the radial component of the null geodesic equation reduces to
\begin{equation}
    \frac{d^2\rho}{d\lambda^2} + V^{(2)}_{\rm eff}(\rho) = \frac{1}{b^2} \ \ \text{with} \ \ 
    V^{(2)}_{\rm eff}(\rho) = \frac{1}{\rho^2 U^4}.
\end{equation}
A physical circular orbit is given by a real root of $dV^{(2)}_{\rm eff}/d\rho=0$. There is no circular orbit if the separation $2d$ is larger than the critical value $2d_{\rm crit}=\frac{2M}{3}\sqrt{\frac{2}{3}}$. On the other hand, for $d<d_{\rm crit}$, we find two circular orbits with the outer radius $\rho_{\rm out}$ and the inner radius $\rho_{\rm in}$. Each radius has an analytical expression \cite{PhysRevD.98.064036, PhysRevD.95.024026}:
\begin{align}
    \frac{\rho_{\rm out}}{M} &= 
    \sqrt{\frac{1}{9}
    \bigg( 1 + 2 \cos\bigg[ \frac{1}{3}\cos^{-1} \bigg( 1 - \frac{27d^2}{M^2} \bigg) \bigg]\bigg)^2-\frac{d^2}{M^2} },
    \label{eq:rho_out}
    \\
    \frac{\rho_{\rm in}}{M} &= 
    \sqrt{\frac{1}{9}
    \bigg( 1 - 2 \sin \bigg[ \frac{\pi}{6} - \frac{1}{3}\cos^{-1} \bigg( 1 - \frac{27d^2}{M^2} \bigg) \bigg]\bigg)^2-\frac{d^2}{M^2} }.
    \label{eq:rho_in}
\end{align}
This implies that the remnant light ring begins to form as early as the separation of the two black holes is $\sim \mathcal{O}(M)$.
We can check the stability in the radial direction by evaluating the sign of the second derivative of the effective potential. 
We then find that the outer one of the two circular orbits is unstable and the inner one is stable. 
To analyze how the closed null orbits on the equatorial plane approach the photon sphere of the extreme Reissnor-Nordstr\"{o}m black hole spacetime in the limit of $d \to 0$, we expand $1-\rho_{\rm out}/M$ in terms of $d/M$ and get
\begin{equation}\label{eq:expanded_rho_out}
    1-\frac{\rho_{\rm out}}{M} = \frac{5}{2} \frac{d^2}{M^2} + \mathcal{O}\left( \frac{d^4}{M^4} \right).
\end{equation}
This shows that the outer circular orbit in the equatorial plane converges to the remnant photon sphere with the deviation of ${\cal O}(d^2/M^2)$.
It means that, at least on the equatorial plane, the common light ring starts to form as early as the separation of the order of $M$ as is shown in Figure \ref{fig:radius_of_outer_circular_orbit}, and it approaches the size of the remnant light ring with the deviation of ${\cal O}(d^2/M^2)$. 
A similar behavior in the deviation is found for the light ring on the meridian plane:
\begin{equation}\label{eq:sol_expanded_r}
    \frac{r(\theta)}{M} = 1+\frac{5(3 + \cos 2\theta) }{12} \frac{d^2}{M^2} + \mathcal{O}\left(\frac{d^4}{M^4}\right),
\end{equation}
where $r(\theta)$ is the radius of the light ring and is defined by the coordinate transformation $\rho = r \sin \theta$ and $z = r \cos \theta$. 
Its detailed derivation can be found in Appendix \ref{sec:app_meridian}.

\section{Conclusion}
\label{sec_conclusion}
In this paper, we proposed a phenomenological model of ringdown amplitude for binary black hole (BBH) mergers. Our model is based on the greybody factor $\Gamma_{lm}(\omega)$, quantifying the absorption nature of a Kerr black hole, and we argue that for the quadrupole moment, the spectral amplitude of gravitational wave (GW) ringdown $|\tilde{h}_{22}(\omega)|$ can be modeled by the greybody factor with $|\tilde{h}_{22}(\omega)| \sim C_{\rm amp} \sqrt{1- \Gamma_{22}(\omega)}$ for $\omega \gtrsim f_{22}$ (see Figure \ref{fig:imr}), where $f_{22}$ is the real part of the fundamental quasinormal (QN) mode frequency. This phenomenological model is independent of another ringdown model, i.e., superposed QN modes, which may involve many fitting parameters and lead to the overfitting. Our model has an advantage in reducing fitting parameters, i.e., it has the remnant mass $M$, spin $j$, and an overall amplitude only. Also, it is known that there is an ambiguity in the start time of ringdown and it propagates to uncertainty in the time-domain data analysis. However, our model works in the frequency domain and the frequency region relevant to merger and ringdown comes at $\omega \gtrsim f_{22} (M,j)$, depending on the remnant parameters, which is less ambiguous compared with the time-domain analysis of ringdown.

To show the validity and limitation of our model, we studied the consistency between our model and the spectral amplitude of GW waveforms provided in the SXS catalog. We confirmed that the exponent of the exponential damping of GW spectral amplitude at high frequencies $\omega \gtrsim f_{22}$ is well consistent with that of the reflectivity $\sqrt{1- \Gamma_{22}(\omega)}$ (see Figures \ref{fig:greybody_multicomp} and \ref{fig:spin_comp}). The two exponents are well-matched with each other, especially for comparable mass ratios.
We also demonstrate the mass-spin inference of the remnant black hole by computing the mismatch ${\cal M}$ between the greybody factor model and the SXS data. We then found that the remnant parameters leading to the least value of ${\cal M}$ are consistent with the true remnant values (see Figure \ref{fig:mismatch}).

Our model is relevant to another model for the merger-ringdown amplitude proposed in Refs. \cite{Khan:2015jqa,Husa:2015iqa}. In our model, the exponential damping in the amplitude is given by the damping in the greybody factor, but their model uses the damping of the fundamental QN mode, $\tau_{lm0}$, with an extra free parameter $\gamma_3 \in [1.25,1.36]$ \cite{Khan:2015jqa,Husa:2015iqa}. It would be an interesting direction to apply our greybody factor model to construct a more accurate phenomenological model with a smaller number of parameters to contribute to the existing model.
The tidal effective one-body post-merger (TEOBPM) model \cite{Damour:2014yha,DelPozzo:2016kmd,Nagar:2018zoe,Nagar:2019wds,Nagar:2020pcj} is one of the sophisticated phenomenological model of ringdown. It models the QN mode-rescaled ringdown waveform with several fitting parameters \cite{Damour:2014yha}. It is an interesting question if one could further reduce the number of fitting parameters in the TEOBPM model by applying the greybody factor or the Green's function of black hole perturbations which has the phase information.

The greybody factor used in our phenomenological model is nothing but the transmissivity of the light ring of the remnant Kerr black hole. In the merger phase of a comparable mass-ratio BBH, the spacetime would be non-linearly disturbed. The light ring of the remnant Kerr black hole should form soon after or even during the merger. Otherwise, it is difficult to interpret how our model works even for comparable mass mergers. We then consider the Majumdar-Papapetrou (MP) solution as a simple analytic model and investigate the relation between the separation distance of two charged black holes $d$ and the size of the light ring.
We found that the light ring on the equatorial plane forms at the critical distance of $d/M = d_{\rm crit}/M \coloneqq \sqrt{2/3}/3 \simeq 0.27$ and the radius of the light ring $r_{\rm e}$ approaches to that of the remnant extremal Reisner-Nordstr\"{o}m black hole $r_{\rm rem}$ with $\Delta \propto d^2$, where $\Delta \coloneqq |r_{\rm e}-r_{\rm rem}|$. A similar behavior is found for the outermost closed null orbit on the meridian plane. Note that in the paper, we have investigated the light ring only on the two planes, equatorial and meridian planes.
It makes sense that the merger-ringdown part of the GW spectrum can be modeled by the greybody factor if the light ring forms at the early stage of the merger phase, similar to the light ring on the equatorial plane of the MP solution, and if the spacetime can be described by the perturbation of the Kerr solution except for the interior of the outermost common horizon.\footnote{We do not expect that the common horizon completely screens the non-linear effects. It is well known that the excitation of quadratic QN modes is important to model the early ringdown precisely (see e.g. Refs.~\cite{Cheung:2022rbm,Mitman:2022qdl}). The greybody factor presented here works at most in the leading-order level.} This statement may be relevant to the analysis performed in Refs. \cite{Nichols:2010qi,Nichols:2011ih}. They demonstrated that black hole perturbation can describe the non-linear aspects of a binary merger accessible to observers far from the collision. Other analysis based on the Effective One-Body method \cite{Buonanno:2000ef} or the Backwards One-Body method \cite{McWilliams:2018ztb} support a scenario implying the early formation of the light ring \cite{McWilliams:2018ztb}.

Our discussion is based on the MP solution to see the formation of the light ring in an analytic way. However, the MP solution is a static solution and has no dynamics in it. What we could shed light on with this model is the relation between the separation distance of two black holes and the formation of the light ring. Defining the light ring in a dynamical spacetime would be challenging but should be important as GW ringdown is sourced by the light ring.

\begin{acknowledgements}
N.~O. is thankful for the valuable discussions with Niayesh Afshordi, Yanbei Chen, and Sizheng Ma.
K.~O.~is supported by JSPS KAKENHI Grant No.~JP23KJ1162. N.~O.~is supported by the Grant-in-Aid for Scientific Research (KAKENHI) project for FY2023 (23K13111) and by the Hakubi project at Kyoto University.
\end{acknowledgements}

\appendix
\section{Extraction of $T^{\rm (GW)}$ from the SXS data}
\label{app_extraction}
We here describe the methodology to extract the value of $T^{\rm (GW)}$  which is the exponent parameter of the exponential damping in the GW spectral amplitude at $\omega \gtrsim f_{22}$. We first perform the Fourier transform with the formula of (\ref{fourier_amplitude}) in the frequency range of $10^{-3} \leq \omega \leq 1$ divided into $10^3$ bins. We then obtain the spectral amplitude as is shown in Figure \ref{fig:greybody_multicomp}. We then perform the non-linear model fitting to fit the function $\tilde{C}_{\rm amp} \exp{(-\omega/T^{\rm (GW)})}$ that has two fitting parameters, $\tilde{C}_{\rm amp}$ and $T^{\rm (GW)}$, with the absolute square of the spectral amplitude in the frequency range of
\begin{equation}
\alpha f_{22} \leq \omega \leq 1.
\end{equation}
The value of the constant $\alpha$ we take is shown in Table \ref{table:alpha}.
\begin{table}[h]
\begin{center}
\begin{tabular}{c c c} 
 \hline
  spin range & cut off parameter ($\alpha$) \\ [0.5ex] 
 \hline\hline
 $0.85\leq j$  & 1.02 \\
 \hline
 $0.65\leq j < 0.85$ & 1.05 \\
 \hline
 $j < 0.65$ & 1.1 \\
 \hline
\end{tabular}
\caption{The cut off parameter $\alpha$ we set in our analysis.}
\label{table:alpha}
\end{center}
\end{table}

\section{Mass-spin extraction from the data with and without the PN correction}
\label{app:mass_spin_noPN}
\begin{figure}[t]
\centering
\includegraphics[width=0.8\linewidth]{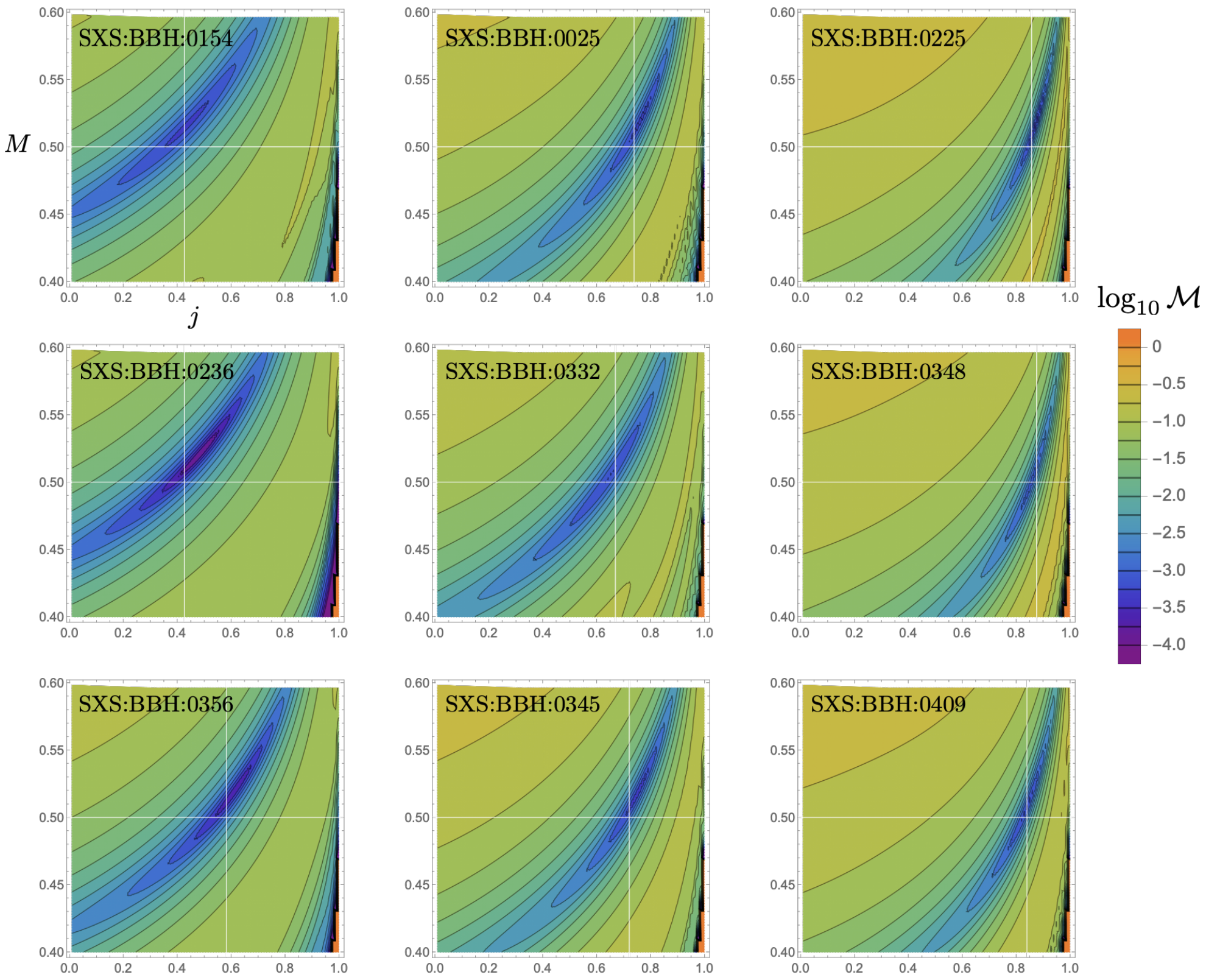}
\caption{Mismatch between our model $C_{\rm amp}{\cal R}$ and the spectral amplitude of the SXS's waveforms. The white lines indicate the true value of the remnant quantities and the remnant mass $M_{\rm true}$ is normalized with $M_{\rm true}=0.5$. We set $\omega_f = 1.7$ and $\alpha = 0.85$.}
\label{fig:085_noPN}
\end{figure}
We here show our analysis of the mass-spin extraction from the SXS data with the greybody factor model. In Sec. \ref{sec_modeling_ringdown}, we demonstrate the mass-spin extraction for three SXS waveforms only. We here apply the greybody-factor analysis to other SXS waveforms. The frequency cut-off are $\omega_i = \alpha \times f_{22}$ and $\omega_f = 1.7$. We here set $\alpha =0.85$ for all waveforms and the results with the greybody factor model without the PN correction are shown in Figure \ref{fig:085_noPN}.

\begin{figure}[t]
\centering
\includegraphics[width=0.8\linewidth]{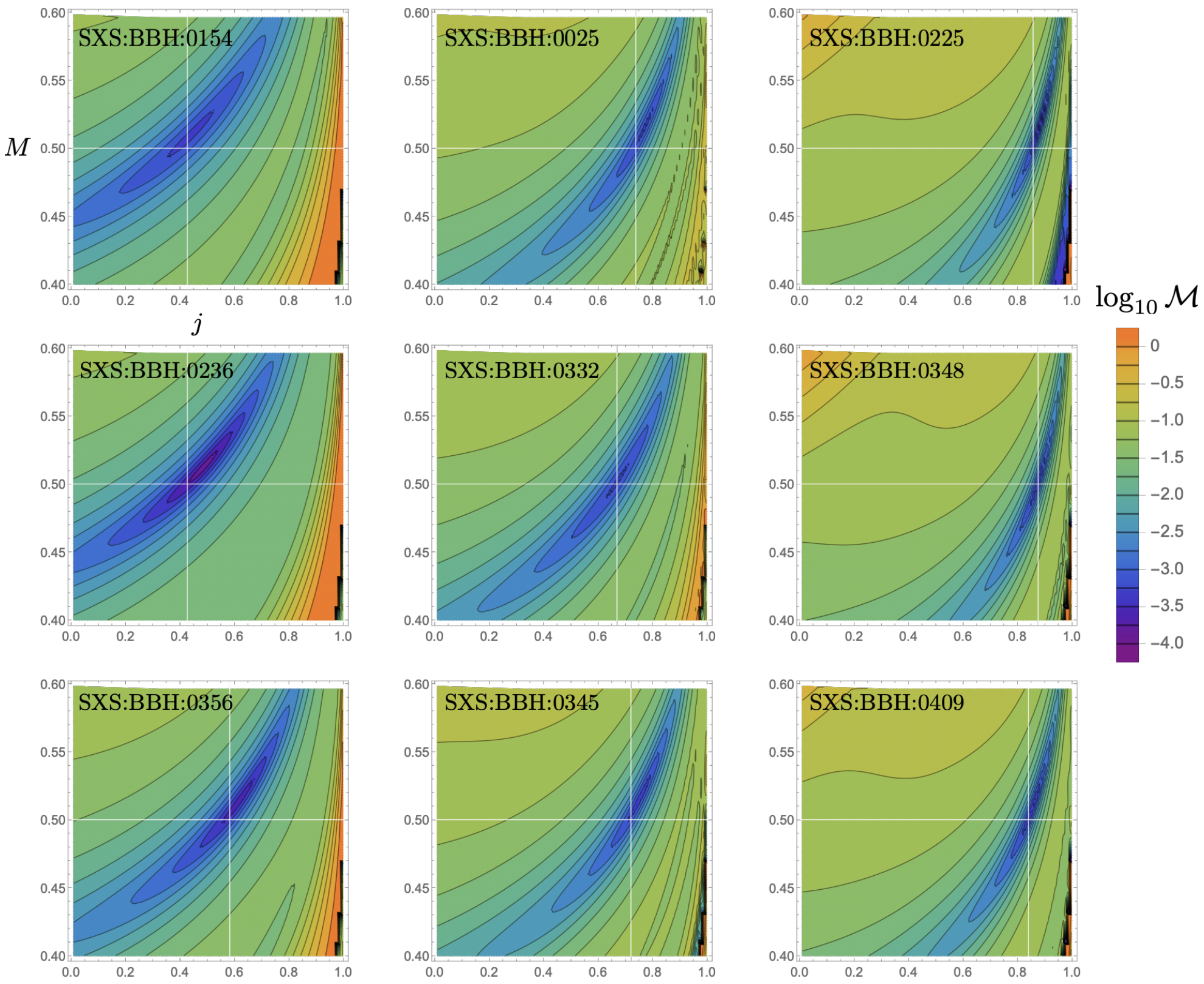}
\caption{Mismatch between the greybody factor model with the PN correction (\ref{greybody_PNexpansion}) and the spectral amplitude of the SXS's waveforms. The white lines indicate the true value of the remnant quantities and the remnant mass $M_{\rm true}$ is normalized with $M_{\rm true}=0.5$. We set $\omega_f = 1.7$ and $\alpha = 0.85$.}
\label{fig:085_PN}
\end{figure}

We also apply our model with the PN correction to the SXS waveforms with the same data range in Figure \ref{fig:085_noPN} and the results are shown in Figure \ref{fig:085_PN}.

\section{The null orbit on the meridian plane in the MP solution}
\label{sec:app_meridian}

\begin{figure}[t]
        \includegraphics[width=0.9\linewidth]{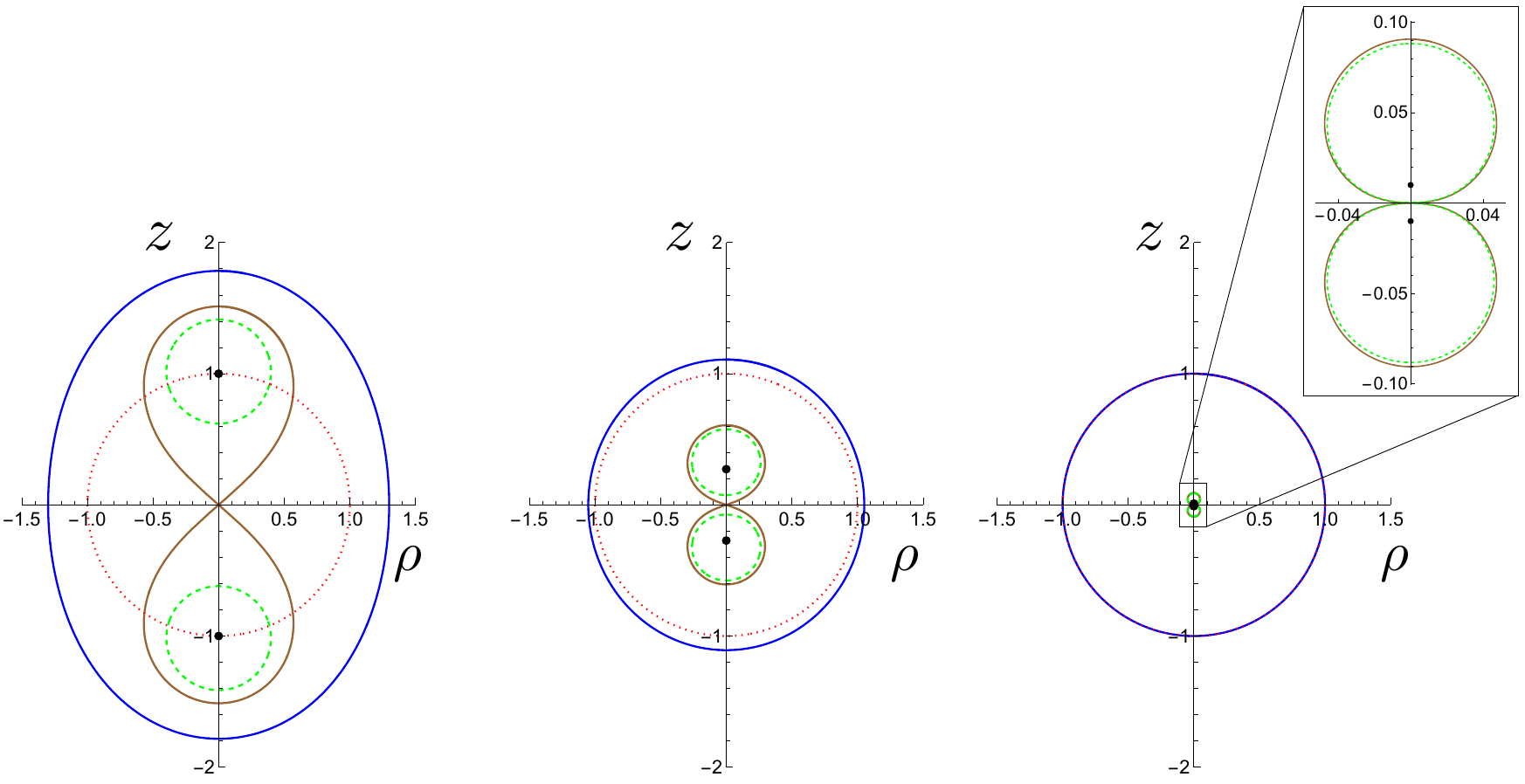}
        \caption{
        The closed null orbits on the meridian plane is shown for $d=M$ (left), $d=d_{\rm crit}$ (center), and $d=10^{-2}M$ (right). As a reference, the circular orbit of the extreme Reissner-Nordstr\"{o}m solution is depicted as a red dashed line. 
        In the limit of $d/M \to 0$, the 8-shaped orbit and the orbits enclosing an individual hole shrink to the origin. However, as seen in the top-right figure, these orbits do not disappear even for smaller values of $d/M$ since the spacetime is static.
        On the other hand, the light ring enclosing the two holes converges to the remnant light ring.
        }
        \label{fig:plot2D_closed_orbits_meridian_plane}
\end{figure}

On the meridian plane, the geodesic equation is not separable, and we need to numerically solve it. 
The null geodesic equations for $\rho$ and $z$ components are written as 
\begin{align}
    &\ddot{\rho} - \frac{ b^2 ( U + \rho \partial_{\rho} U )}{\rho^3 U^5} + \frac{2 \dot{\rho} \dot{z} \partial_{z} U- (1 +\dot{z}^2 - \dot{\rho}^2 ) \partial_{\rho} U } {U}=0,
    \\
    &\ddot{z} - \frac{ b^2 \partial_{z} U}{\rho^2 U^5} + \frac{2 \dot{\rho} \dot{z} \partial_{\rho} U - (1 -\dot{z}^2 +\dot{\rho}^2) \partial_{z} U } {U}=0.
\end{align}
where $b$ is the impact parameter defined by $b=L/E$, and the dot means the derivative with respect to the affine parameter $\lambda$.
To consider closed null geodesics on the meridian plane, we choose the impact parameter as $b=0$ since $L$ is the angular momentum associated with the Killing vectors and written as $L = \rho^2 U^2 \dot{\phi}$.
Thus, we consider the following geodesic equation to find a closed null orbit,
\begin{align}
    &\ddot{\rho} + \frac{2 \dot{\rho} \dot{z} \partial_{z} U- (1 +\dot{z}^2 - \dot{\rho}^2 ) \partial_{\rho} U } {U}=0,
    \label{eq:geodesics_rho}
    \\
    &\ddot{z} + \frac{2 \dot{\rho} \dot{z} \partial_{\rho} U - (1 -\dot{z}^2 +\dot{\rho}^2) \partial_{z} U } {U}=0.
    \label{eq:geodesics_z}
\end{align}
In addition to these equations, we have the null condition:
\begin{equation}\label{eq:null_condition_rho_z}
    \dot{\rho}^2 + \dot{z}^2 = 1.
\end{equation}
If we start to solve the geodesic equations from the $z$-axis at $\lambda=0$, the initial conditions are chosen as $(\rho, z)|_{\lambda=0}=(0, z_0)$, $\dot{z}(0)=0$ where $z_0$ is a coordinate value of the $z$-axis. The initial condition for the derivative of $\rho$ is determined by the null condition Eq. (\ref{eq:null_condition_rho_z}), and $\dot{\rho}(0)=1$ is obtained.
To obtain the closed orbit, we solve the geodesic equations by using the shooting method with the parameter $z_0$. As a result, three types of closed orbits are obtained: the orbit enclosing the two black holes, the 8-shaped orbit, and the orbit enclosing the individual black hole \cite{doi:10.1098/rspa.1989.0010, Shipley_2016, PhysRevD.98.064036}. Notice that each orbit always exists and is unstable as far as we have numerically confirmed.
Figure \ref{fig:plot2D_closed_orbits_meridian_plane} shows several examples of the three closed orbits. The outermost closed orbit (blue solid) converges to the circular orbit on the photon sphere in the remnant extreme Reissner-Nordstr\"{o}m spacetime (red dotted) and the others (brown solid and green dashed) converge to the origin as the separation distance $d$ decreases. Hence, the outermost closed orbit is relevant to the remnant photon sphere.  
To analyze how the outermost closed null orbit on the meridian plane approaches the circular orbit in the extreme Reissnor-Nordstr\"{o}m black hole spacetime, we focus on null geodesics on the hypersurface described by $\phi={\rm const.}$ Hereafter, this hypersurface is denoted by $\Sigma_{\phi}$. The induced metric on $\Sigma_{\phi}$ is given by
\begin{equation}\label{eq:MP_inducedmetric}
    ds^2 = -U^{-2}dt^2+ U^2(d\rho^2+dz^2).
\end{equation}
To study the outermost closed null orbit, we adopt the spherical coordinate $(t, r, \theta)$ through the coordinate transformation on $\Sigma_{\phi}$: $\rho = r \sin \theta$ and $z = r \cos \theta$. Then, the induced metric is written as
\begin{equation}\label{eq:MP_inducedmetric2}
    ds^2 = -\tilde{U}^{-2}(r, \theta) dt^2+ \tilde{U}^2(r, \theta) (dr^2+r^2 d\theta^2),
\end{equation}
where 
\begin{equation}
    \tilde{U}(r, \theta) \coloneqq  1 + \frac{M}{2 \sqrt{r^2 - 2\, d\, r \cos \theta + d^2}} + \frac{M}{2 \sqrt{r^2 + 2\, d\, r \cos \theta + d^2}}.
\end{equation}
The geodesic equations on the hypersurface $\Sigma_{\phi}$ are given by
\begin{align}
    &\ddot{r} - \frac{1}{\tilde{U}} 
        \Big[ r \dot{\theta}^2 \tilde{U} - 2 \dot{r} \dot{\theta} \tilde{U}_{,\theta} +  ( 1 -\dot{r}^2 + r^2 \dot{\theta}^2 ) \tilde{U}_{,r}
        \Big] = 0,
    \\
    &\ddot{\theta} - \frac{1}{r^2 \tilde{U}} 
        \Big[ -2 r \dot{r} \dot{\theta} \tilde{U} - 2 r^2 \dot{r} \dot{\theta} \tilde{U}_{,r} +  ( 1 +\dot{r}^2 - r^2 \dot{\theta}^2 ) \tilde{U}_{,\theta}
        \Big] = 0,
\end{align}
and the null condition is given by
\begin{equation}\label{eq:null_condition_r_theta}
    \dot{r}^2 + r^2 \dot{\theta}^2 = 1.
\end{equation}
Since we focus on the outermost closed null orbit, it is useful to use the coordinate $\theta$ as a parameter of the orbit instead of the affine parameter $\lambda$. 
\begin{figure}[t]
        \includegraphics[width=0.45\linewidth]{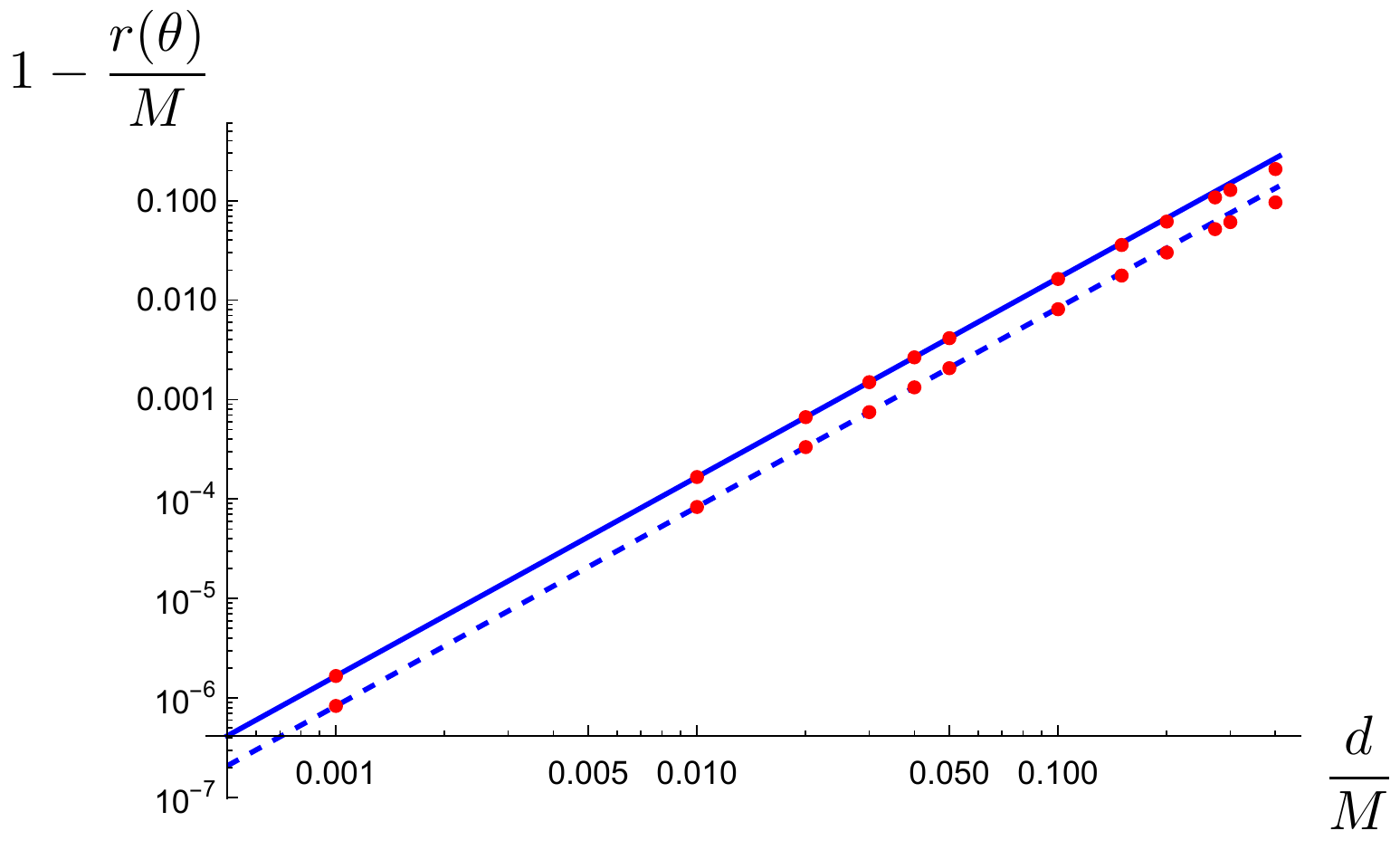}
        \caption{The difference between the size of the photon sphere on the meridian plane for $d > 0$ and that for $d=0$ (i.e. Reissnor-Nordstr\"{o}m black hole spacetime), quantified by $1-r(\theta)/M$, is shown with respect to the separation $d/M$. We show the approximated values $1-r(\theta)/M \simeq (5/12) (3+\cos 2\theta) (d^2/M^2)$ for $\theta = 0$ (solid) and $\theta = \pi/2$ (dashed). The red markers show the values obtained numerically which agrees with the approximated formula in Eq. (\ref{eq:sol_expanded_r_app}).}
        \label{fig:LogPlot2D_meridian_plane}
\end{figure}
After rewriting the above equations, we obtain the following equation:
\begin{equation}\label{eq:geodesic_r_theta2}
    r'' - \frac{2 r'^2}{r} - r 
    + 2(r^2+r'^2) \left[ \frac{r'}{r^2} \frac{U_{,\theta} }{U} 
    - \frac{U_{,r}}{U} \right] 
   = 0,
\end{equation}
where the prime indicates the derivative with respect to $\theta$. The relation between $\lambda$ and $\theta$ is determined by the null condition Eq. (\ref{eq:null_condition_r_theta}). 
Obtaining the analytical solution of the closed orbit from the geodesic equation is difficult in general, but we can iteratively solve the geodesic equation Eq. (\ref{eq:geodesic_r_theta2}) for the sufficiently small coordinate distance $d/M$. 
Since we know that the outermost closed orbit approaches the circular orbit with the radius $M$ from the numerical result, we can expand the function $r(\theta)$ by using dimensionless functions $\tilde{r}_{i}(\theta)~(i=1, 2, 3)$ as
\begin{equation}\label{eq:expansion_r}
    \frac{r(\theta)}{M} = 1 + \tilde{r}_{1}(\theta) \frac{d}{M} + \tilde{r}_{2}(\theta) \frac{d^2}{M^2} + \tilde{r}_{3}(\theta) \frac{d^3}{M^3} + \mathcal{O}\left(\frac{d^4}{M^4}\right).
\end{equation}
Substituting Eq. (\ref{eq:expansion_r}) into the geodesic equation Eq. (\ref{eq:geodesic_r_theta2}), we obtain differential equations at each order of $d/M$. 
At the first order and the third order, we obtain the solutions $\tilde{r}_1(\theta)=\tilde{r}_3(\theta)=0$ by imposing the periodicity of a closed orbit, i.e., $r(0)=r(2\pi)$.
At the second order, the above equation Eq. (\ref{eq:geodesic_r_theta2}) can be expanded as
\begin{equation}\label{eq:geodesic_r_theta_expansion}
    \frac{d^2 \tilde{r}_2}{d \theta^2} - \frac{\tilde{r}_2}{2} - \frac{ 5(1-3 \cos^2 \theta) }{4} = 0.
\end{equation}
This can be solved analytically, and the solution is given by,
\begin{equation}\label{eq:sol_expanded_r_2nd}
    \tilde{r}_2 = \frac{ 5(3 + \cos 2\theta) }{12}.
\end{equation}
As a result, the radius $r(\theta)$ of the outermost closed null orbit is given by
\begin{equation}\label{eq:sol_expanded_r_app}
    \frac{r(\theta)}{M} = 1 + \frac{5(3 + \cos 2\theta) }{12} \frac{d^2}{M^2} + \mathcal{O}\left(\frac{d^4}{M^4}\right),
\end{equation}
for $d/M \ll 1$. This shows that the outermost circular orbit enclosing two black holes on the meridian plane converges to the remnant photon sphere of the extreme Reissnor-Nordstr\"{o}m spacetime with the deviation of ${\cal O}(d^2/M^2)$. As shown in Figure \ref{fig:LogPlot2D_meridian_plane}, the numerical results of the radius agree with the approximated solution Eq. (\ref{eq:sol_expanded_r_app}) in the limit of $d/M \to 0$. 
This does not change our statement concluded from the analysis on the equatorial plane, i.e., that the common light ring may form as early as the separation of the two black holes is $\sim {\cal O}(M)$.

\end{document}